\documentclass[a4paper,11pt]{article}
\pdfoutput=1 

\usepackage{jheppub} 

\usepackage{amsmath,amssymb}
\usepackage{slashed}

\def\d{\partial}
\def\p{{\mathbf p}}
\def\q{{\mathbf q}}
\def\k{{\mathbf k}}
\def\x{{\mathbf x}}
\def\y{{\mathbf y}}
\def\r{{\mathbf r}}
\def\b{{\mathbf b}}
\def\z{{\mathbf z}}
\def\v{{\mathbf v}}
\def\u{{\mathbf u}}
\def\w{{\mathbf w}}
\def\ux{{\underline{x}}}
\def\uy{{\underline{y}}}
\def\uz{{\underline{z}}}
\def\uk{{\underline{k}}}

\def\uq{{\underline{q}}}
\def\d{\partial}
\def\p{{\mathbf p}}
\def\q{{\mathbf q}}
\def\k{{\mathbf k}}
\def\x{{\mathbf x}}
\def\y{{\mathbf y}}
\def\r{{\mathbf r}}
\def\z{{\mathbf z}}
\def\v{{\mathbf v}}
\def\u{{\mathbf u}}
\def\ux{{\underline{x}}}
\def\uy{{\underline{y}}}
\def\uz{{\underline{z}}}
\def\uk{{\underline{k}}}

\def\uq{{\underline{q}}}
\newcommand \nn {\nonumber}

\newcommand{\be}{\begin{equation}}
\newcommand{\ee}{\end{equation}}
\newcommand{\ba}{\begin{eqnarray}}
\newcommand{\ea}{\end{eqnarray}}
\newcommand{\ban}{\begin{eqnarray*}}
\newcommand{\ean}{\end{eqnarray*}}

\title{Quarks at next-to-eikonal accuracy in the CGC I:
Forward quark-nucleus scattering}

\author[a]{Tolga Altinoluk,}
\author[a]{Guillaume Beuf,}
\author[a]{Alina Czajka}
\author[a]{and Arantxa Tymowska}

\affiliation[a]{Theoretical Physics Division, National Centre for Nuclear Research, \\
Pasteura 7, Warsaw 02-093, Poland}

\emailAdd{tolga.altinoluk@ncbj.gov.pl}
\emailAdd{guillaume.beuf@ncbj.gov.pl}
\emailAdd{alina.czajka@ncbj.gov.pl}
\emailAdd{arantxa.tymowska@ncbj.gov.pl}

\abstract{ We derive Next-to-Eikonal (NEik) corrections to the background quark propagator, which stem from (i) considering a finite longitudinal width target instead of an infinitely thin shockwave and (ii) including the interaction of the quark with the transverse components of the background field. These two different corrections to the eikonal approximation combine together and provide a gauge covariant expression for the quark propagator at NEik accuracy. We then apply our results to
quark (or antiquark) scattering on a nucleus at NEik accuracy, considering both unpolarized cross section and helicity asymmetry. }

\begin{document}
\maketitle

\section{Introduction}
\label{sec:intro}

High energy hadronic scattering processes in the dilute-dense systems, such as proton-nucleus (pA) collisions, are described within the effective theory known as the Color Glass Condensate (CGC) (see \cite{Gelis:2010nm} for a review and references there in).  In the CGC effective theory, the dilute projectile is described by its color charge density while the dense target is described by a strong background field. One of the key approximations that is routinely adopted in the standard CGC calculations is known as the eikonal approximation.

High scattering energies in dilute-dense systems are achieved by boosting the dilute projectile and the dense target from their rest frames to the scattering energies. From the target point of view, the eikonal approximation in such a case amounts to employing the following three assumptions: (i) the highly boosted background field that describes the target is localized in the longitudinal direction (around $x^+=0$) due to Lorentz contraction, (ii) the background field of the target is given by its leading component (which in our setup corresponds to the ``$-$" component) whereas the subleading components in terms of  the Lorentz boost factor (the transverse and ``$+$" components) are neglected, and (iii)  the background field of the target is assumed to be independent of the $x^-$ due to Lorentz time dilation, which amounts to neglecting the dynamics of the target. In this case, which is also known as the shockwave approximation, the background field of the target is described by
\be
{\cal A}_a^{\mu}(x^-,x^+,\x)\approx\delta^{\mu -}\delta(x^+){\cal A}_a^-(\x)
\ee
Within the eikonal approximation, the projectile partons undergo multiple scatterings while propagating through the dense target and this information is encoded via Wilson lines that are defined as path ordered exponentials of the leading component of the background field of the target. In practice, the eikonal approximation amounts to describing the scattering processes by the contributions leading in energy and neglecting all the power-suppressed contributions.

While the eikonal approximation is  a very successful and powerful tool to describe the asymptotically high energy scatterings, in reality the scattering energies that are reached in the experiments are limited. Therefore, understanding the applicability region of the eikonal approximation and computing the corrections to this approximation is of key importance for a precise comparison of the theoretical calculations to the experimental data. Even though for the scattering energies achieved at the Large Hadron Collider (LHC) the eikonal approximation might still be reliable, subeikonal corrections are expected to play a more significant role at the Relativistic Heavy Ion Collider (RHIC) energies. Moreover, the planned Electron-Ion Collider (EIC) in the USA will not probe extremely high energies, thus  subeikonal corrections might be also sizable in EIC phenomenology.

One of the observables that is frequently studied in the CGC framework within the standard eikonal approximation is particle production. For production in the mid-rapidity region, both the projectile and the target are treated within the CGC framework and this approach is known as the $k_t$- factorization (see \cite{Kovchegov:1998bi, Dumitru:2001ux, Kovchegov:2001sc, Kharzeev:2003wz, Blaizot:2004wu} for mid-rapidity gluon production, see \cite{Gelis:2003vh, Blaizot:2004wv} for quark production). On the other hand, hybrid factorization \cite{Dumitru:2002qt, Dumitru:2005gt} is the state-of-the-art framework for particle production at forward rapidity.  In this setup, the dilute projectile is treated in the collinear framework while the dense target is treated within the standard CGC. During the last decade, a vast amount of effort has been devoted to the theoretical computations of the next-to-leading order (NLO) corrections in strong coupling constant \cite{Altinoluk:2011qy, Chirilli:2011km, Chirilli:2012jd, Stasto:2013cha, Kang:2014lha, Stasto:2014sea, Altinoluk:2014eka, Watanabe:2015tja, Ducloue:2017dit, Iancu:2016vyg, Liu:2019iml, Liu:2020mpy} and its numerical applications \cite{Ducloue:2016shw, Ducloue:2017mpb} in forward hadron production. Moreover, within a hybrid set up, forward heavy quark production \cite{Altinoluk:2015vax, Marquet:2017xwy}, forward multi-jet production \cite{Iancu:2018hwa, Iancu:2020mos} and its relation to the transverse momentum dependent distributions \cite{vanHameren:2014lna, Kotko:2015ura, vanHameren:2016ftb, Altinoluk:2018uax, Altinoluk:2018byz, Altinoluk:2020qet} are also studied.

Since it provides a clean environment to probe gluon saturation effects, deep inelastic scattering (DIS) on a dense target is another observable studied within the CGC framework. Most of the DIS related observables are computed in dipole factorization which was first introduced in \cite{Nikolaev:1990ja} and it has a very clear physics interpretation: the virtual photon emitted from the lepton splits into a quark-antiquark pair which then scatters on the dense target. The splitting of the virtual photon can be computed perturbatively,  while the dipole scattering amplitude which accounts for the interaction of the quark-antiquark pair with the dense target is described within the CGC framework. There has been a lot of effort to compute DIS related observables at NLO accuracy. Inclusive DIS for massless quarks \cite{Balitsky:2010ze, Balitsky:2012bs, Beuf:2011xd, Beuf:2016wdz, Beuf:2017bpd, Hanninen:2017ddy}, diffractive dijet production in DIS \cite{Boussarie:2016ogo}, exclusive light vector meson production in DIS \cite{Boussarie:2016bkq}, photon + dijet production in DIS \cite{Roy:2019hwr} have already been computed at NLO accuracy.

As mentioned previously, for a precise comparison of the CGC-based theoretical studies to the experimental data, it is very important to consider the corrections to the eikonal approximation especially for processes with not extremely high collision energies. This idea triggered  a lot of activities to go beyond the eikonal accuracy within the CGC framework. In \cite{Altinoluk:2014oxa, Altinoluk:2015gia}, the eikonal approximation was relaxed by considering a finite width target instead of an infinitely thin shockwave and a systematic approach was introduced to compute the finite-width corrections to the gluon propagator traversing a dense target at Next-to-Next-to-Eikonal (NNEik) accuracy. These corrections are associated with the transverse motion of the gluon during its interaction with the target.  The results were then applied to the production of single inclusive gluon and various spin asymmetries in proton-nucleus collisions at mid-rapidity, and corrections to the eikonal approximation for these observables were computed. Later in \cite{Altinoluk:2015xuy, Agostini:2019avp}, it was shown that in the weak field limit, the finite-width corrections can be interpreted as sub-eikonal corrections to the standard Lipatov vertex and multi gluon production cross section was computed with non-eikonal corrections for proton-proton collisions at mid-rapidity. The effects of non-eikonal corrections (due to finite width of the target) on two-particle correlations were studied in \cite{Agostini:2019hkj}. Apart from the aforementioned studies that focus on the finite-width corrections to gluon production at mid-rapidity to go beyond the eikonal accuracy, in \cite{Kovchegov:2015pbl, Kovchegov:2016weo, Kovchegov:2016zex, Kovchegov:2017jxc, Kovchegov:2017lsr, Kovchegov:2018znm, Kovchegov:2020hgb, Adamiak:2021ppq} quark and gluon helicity distributions were derived at Next-to-Eikonal (NEik) order. Recently, helicity-dependent generalization of the non-linear rapidity evolution that goes beyond eikonal accuracy has been derived in \cite{Cougoulic:2019aja, Cougoulic:2020tbc}. In \cite{Chirilli:2018kkw}, the sub-eikonal corrections to both quark and gluon propagators were derived within the high-energy operator product expansion (OPE) formalism. On the other hand, the studies beyond eikonal accuracy have been also performed in the context of transverse momentum dependent parton distributions (TMDs). The rapidity evolution equations for gluon TMDs that interpolate between high and moderate values of energy, thus accounting for sub-eikonal effects, have been derived in \cite{Balitsky:2015qba, Balitsky:2016dgz, Balitsky:2017flc, Balitsky:2017gis, Balitsky:2019ayf, Balitsky:2020jzt}. A similar idea is also proposed in \cite{Boussarie:2020fpb} where a formulation for unintegrated gluon distributions that interpolates between the small and moderate values of energy has been studied. A different approach based on exchanging longitudinal momentum during the scattering to go beyond eikonal accuracy is being pursued in \cite{Jalilian-Marian:2017ttv, Jalilian-Marian:2018iui, Jalilian-Marian:2019kaf}. Also, NEik corrections were computed in the context of soft gluon exponentiation for example in \cite{Laenen:2008ux, Laenen:2008gt, Laenen:2010uz}.

In this work, we focus on the quark propagator in a gluon background and compute the full NEik corrections to it. These NEik corrections that we compute in this work originate from relaxing two different assumptions of the eikonal approximation. On the one hand, we consider a finite width target instead of an infinitely thin shockwave, therefore we account for the NEik corrections that are due to finite longitudinal width. These corrections are the analogue of the ones computed in \cite{Altinoluk:2014oxa} for the gluon propagator and they are associated with the transverse motion of the projectile parton during its interaction with the target. On the other hand, as discussed previously, in the eikonal approximation the target background field is defined by its leading component which in our setup is given by the "$-$" component. Here, we also relax this approximation, include the transverse components of the target background field and compute the associated NEik corrections. By combining these two effects, we are able to write the full NEik quark propagator in a compact form which is gauge covariant. We show that helicity dependent part of the quark propagator starts at NEik  order while the helicity independent part starts at eikonal order.

The plan for the rest of the paper is as follows. In Section \ref{sec:quark_prop}, we present the derivation of the quark propagator at NEik accuracy by including both the transverse motion effects and the effects of transverse component of the background field. In Section \ref{sec:cross_section}, we apply our results to forward quark-nucleus scattering  where we first compute the quark scattering amplitude at NEik accuracy by using the LSZ reduction formula, and then use this amplitude to compute both the unpolarized cross section and quark helicity asymmetry at partonic level. Finally, in Section \ref{sec:Discussions}, we summarize our results and discuss future plans. The details of the derivation of the transverse motion effects at NEik accuracy are presented in Appendix \ref{appendix}. In Appendix \ref{Sec:Comparison}, we compare our results for the quark propagator at NEik accuracy with the ones obtained in \cite{Chirilli:2018kkw}.{\footnote{ Appendix \ref{Sec:Comparison} has been added in the second version of the present manuscript, following the release of the paper \cite{Chirilli:2021lif}, in which multiple apparent differences were pointed out between the results of the present manuscript and the ones of Ref. \cite{Chirilli:2018kkw}.}} The study of forward antiquark-nucleus scattering can be found in Appendix \ref{Sec:antiquark-observables}.

\section{Quark propagator through a shockwave at NEik accuracy}
\label{sec:quark_prop}

\subsection{Setup}

This section is devoted to the study of the quark propagator $S_F(x,y)_{\beta\alpha}$ (with $\beta, \alpha$ the fundamental color indices) in a classical background gluon field $\mathcal{A}^a_{\mu}(x)$. It can be written as a sum of the vacuum contribution and a medium correction as
\begin{align}
S_F(x,y)_{\beta\alpha} &= S_{0,F}(x,y)_{\beta\alpha} +  \delta S_F(x,y)_{\beta\alpha}
\, ,
\end{align}
with the vacuum quark propagator
\begin{align}
{S}_{0,F}(x,y)_{\beta\alpha} &= \left(\mathbf{1}\right)_{\alpha\beta}\, \int \frac{d^4k}{(2\pi)^4}\:
e^{-i k\cdot (x-y)}\;  \frac{i(\slashed{k}+m)}{[k^2\!-\!m^2+i\epsilon]}
\label{free_q_prop_Fourier}
\, .
\end{align}
The chosen convention for Fourier transform is
\begin{align}
S_F(x,y)_{\beta\alpha} &= \int \frac{d^4q}{(2\pi)^4} \int \frac{d^4k}{(2\pi)^4}\;
e^{-i x\cdot {q}}\; e^{+i y\cdot {k}}\;  \tilde{S}_F(q,k)_{\beta\alpha}
\label{q_prop_Fourier}
\, .
\end{align}

We consider the case of a classical background field (with no gauge condition imposed a priori) which is non-zero only in a bounded support, in particular along the light-cone direction $x^+$.\footnote{Our conventions and notations for spacetime quantities are as follows. The chosen signature for the metric is $(+,-,-,-)$. For a Minkowski 4-vector $x$, with component indices $\mu$, $\nu$, \dots, one writes its transverse components (with indices $i$, $j$, \dots) as a Euclidian transverse vector $\x$. The light-cone components are defined as $x^{\pm}= (x^0\pm x^3)/\sqrt{2}$. We will also use the notation $\ux$ for the restriction of $x$ to the $+$ and transverse directions, $\ux=(x^+, \x)$. The same dot is used for Minkowski and Euclidian products, hence $\x\!\cdot\!\y = x^1 y^1 + x^2 y^2$, $x\!\cdot\! y = x^+ y^- + x^- y^+ - \x\!\cdot\!\y$, and $\ux\!\cdot\! y = x^+ y^- - \x\!\cdot\!\y$.
Finally, the notation $\check{k}$ is used for a momentum 4-vector with the same $+$ and transverse components as $k$, but with $\check{k}^-$ chosen such that $\check{k}$ is on-shell, $\check{k}^-=(\k^2+m^2)/(2k^+)$ for a quark of mass $m$ or $\check{k}^-=\k^2/(2k^+)$ for a gluon.}
The high-energy regime of interest here corresponds to the case in which the background field is highly boosted along the $x^-$ direction, with a large Lorentz factor $\gamma \gg 1$. There is then a strong hierarchy between the components of the background field:
\begin{align}
& \mathcal{A}^- =O(\gamma) \gg \mathcal{A}_j =O(1) \gg  \mathcal{A}^+ =O(1/\gamma)
\, .
\end{align}
The support of the background field along the $x^+$ direction is chosen to be $[-L^+/2, L^+/2]$, without loss of generality, with $L^+ =O(1/\gamma)$ small due to Lorentz contraction. Each vertex of interaction of quark with the background field comes with an integration in $x^+$ over $[-L^+/2, L^+/2]$, and thus brings a suppression factor $1/\gamma$. In the case of a $\mathcal{A}^-$ insertion, the enhancement of the field component itself compensates the suppression due to the integration. Hence, multiple interactions in $\mathcal{A}^-$ have to be resummed to all orders for $S_F(x,y)$ already at leading power in the boost factor $\gamma$, corresponding to the Eikonal approximation.

By contrast, an insertion of the transverse components $\mathcal{A}_j$ in $S_F(x,y)$ is a priori a contribution suppressed as $1/\gamma$, corresponding to NEik accuracy, and an insertion of $\mathcal{A}^+$ in $S_F(x,y)$ is a priori suppressed as $1/\gamma^2$, corresponding to NNEik  accuracy. In this study, we restrict ourselves to NEik accuracy, so that the $\mathcal{A}^+$ component can be safely ignored.

Finally, the $x^-$ dependence of the background field is weak due to Lorentz time dilation. Accordingly, in the Eikonal approximation, the background field is assumed to be independent of $x^-$. By gauge invariance, effects of $x^-$ dependence should be tied (within covariant derivatives or field strength components) to the $\mathcal{A}_-=\mathcal{A}^+$ component of the background field, and are thus expected to be relevant only at NNEik  accuracy as well. For that reason, the background field will still be assumed to be $x^-$ independent in this study, $\mathcal{A}^a_{\mu}(x)=\mathcal{A}^a_{\mu}(x^+,\x)\equiv\mathcal{A}^a_{\mu}(\ux)$ in our notations.

\subsection{Quark propagator in a pure $\mathcal{A}^-$ background}

Let us focus first in this section on the quark propagator in a background field with only the $\mathcal{A}^-$ component, since interactions with it have to be resummed to all orders, before including the transverse components $\mathcal{A}_j$ of the background field as a perturbation in the following sections.

\subsubsection{General expression in momentum space}

\begin{figure}[!t]
\centering
\includegraphics[scale=0.5]{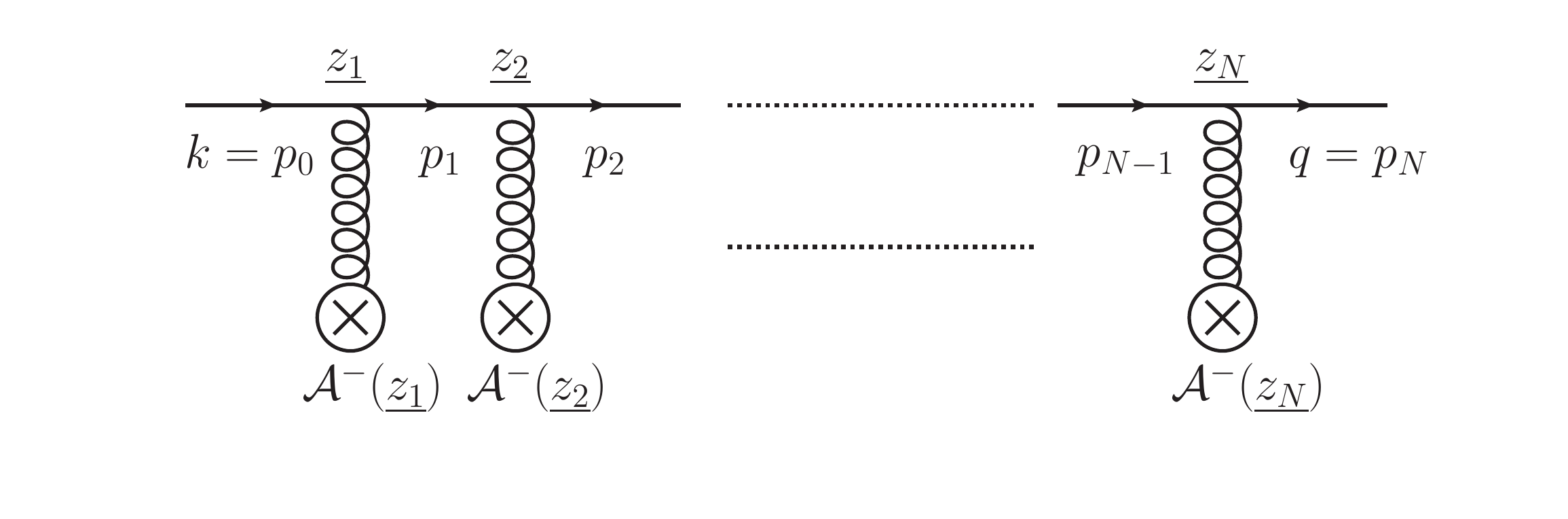}
\caption{Multiple scattering of a quark on a pure $\mathcal{A}^-$ background field.}
\label{fig:multi_A_min}
\end{figure}

In the case of a pure $\mathcal{A}^-$ background field, the medium correction to the quark propagator in momentum space is obtained by summing multiple interaction diagrams of the type shown on Fig.~\ref{fig:multi_A_min}. That sum can be written as\footnote{We use the convention ${{D}_{x^{\mu}}} \equiv  {\d_{x^{\mu}}} +igt\!\cdot\!{A}_{\mu}(x)$ for the covariant derivative in QCD, so that the quark-gluon vertex is $-ig \gamma^{\mu} [t^a]_{\beta \alpha}$.}
\begin{align}
&\delta \tilde{S}_F(q,k)_{\beta\alpha} \bigg|_{\textrm{pure }\mathcal{A}^{-}}
=  \sum_{N=1}^{+\infty} \int \left[\prod_{n=1}^{N-1}  \frac{d^4 p_n}{(2\pi)^4}\right]
\int\left[\prod_{n=1}^{N}  d^3\underline{z_n}\: e^{i \underline{z_n}\cdot (p_n-p_{n-1})}\: 2\pi\delta(p_n^+\!-\!p_{n-1}^+)  \right]
\nn\\
& \times\:
\left\{ \mathcal{P}_n \prod_{n=1}^{N} \left[-igt\!\cdot\!\mathcal{A}^-(\underline{z_n})\right]
\right\}_{\beta\alpha}\:
\frac{i(\slashed{q}\!+\!m)}{[q^2\!-\!m^2+i\epsilon]}
\left\{ \mathcal{P}_n \prod_{n=0}^{N-1} \left[\gamma^+ \frac{i(\slashed{p}_n\!+\!m)}{[{p_n}^2\!-\!m^2+i\epsilon]}
\right]
\right\}
\, ,
\label{q_prop_pure_A_minus_Fourier_1}
\end{align}
where $p_0\equiv k$ and $p_N\equiv q$. In Eq.~\eqref{q_prop_pure_A_minus_Fourier_1}, the symbol $\mathcal{P}_n$ indicates the ordering of the matrix factors with increasing $n$ from right to left, for color matrices in the first bracket and Dirac matrices in the second bracket.
 Due to the fact that $g^{++}=0$ in light-cone coordinates, one has the identity $\gamma^+ \gamma^+=0$, and thus
\begin{align}
\gamma^+(\slashed{p}_n\!+\!m)\gamma^+ = \{\gamma^+,\slashed{p}_n\}\, \gamma^+ = (2p_n^+)\, \gamma^+\, ,
\end{align}
which can be used to simplify the numerator in Eq.~\eqref{q_prop_pure_A_minus_Fourier_1}. Then, for each internal free fermion propagator in Eq.~\eqref{q_prop_pure_A_minus_Fourier_1}, the integration over $p_n^-$ can be performed as
\begin{align}
&\int \frac{dp_n^-}{2\pi}\: e^{-ip_n^-(z_{n+1}^+-z_{n}^+)}\:    \frac{(2p_n^+)i}{[{p_n}^2\!-\!m^2+i\epsilon]}
\nn\\
=& \left\{ \theta(p_n^+) \theta(z_{n+1}^+\!-\!z_{n}^+)-\theta(-p_n^+) \theta(z_{n}^+\!-\!z_{n+1}^+) \right\}\;
e^{-i\check{p}_n^-(z_{n+1}^+-z_{n}^+)}
\, .
\label{p_minus_integration}
\end{align}
Eq.~\eqref{q_prop_pure_A_minus_Fourier_1} then becomes\footnote{Note that, thanks to the intermediate $\gamma^+$, terms like $q^- \gamma^+$ and $k^- \gamma^+$ from the numerators of the external free propagators drop, so that no $q^-$ or $k^-$ survive in the numerator (but only in the denominators and in the phase factors). In order to emphasize this fact, $\slashed{q}$ and $\slashed{k}$ have been replaced by their on-shell counterparts in Eq.~\eqref{q_prop_pure_A_minus_Fourier_2}, $\slashed{\check{q}}$ and $\slashed{\check{k}}$ respectively.}
\begin{align}
\delta \tilde{S}_F(q,k)_{\beta\alpha}& \bigg|_{\textrm{pure }\mathcal{A}^{-}}
= \, 2\pi\delta(q^+\!-\!k^+)\, \frac{i(\slashed{\check{q}}\!+\!m)}{[q^2\!-\!m^2+i\epsilon]}\gamma^+\frac{i(\slashed{\check{k}}\!+\!m)}{[k^2\!-\!m^2+i\epsilon]}
\sum_{N=1}^{+\infty}\int\left[\prod_{n=1}^{N}  d^3\underline{z_n}\right]\;
\nn\\
& \times\:
\left\{ \theta(k^+) \left(\prod_{n=1}^{N-1}\theta(z_{n+1}^+\!-\!z_{n}^+)\right)
+(-1)^{N-1}\,\theta(-k^+) \left(\prod_{n=1}^{N-1}\theta(z_{n}^+\!-\!z_{n+1}^+)\right) \right\}\;
\nn\\
& \times\:
e^{i \underline{z_N}\cdot q}\, e^{-i \underline{z_1}\cdot k}\:
\left\{ \mathcal{P}_n \prod_{n=1}^{N} \left[-igt\!\cdot\!\mathcal{A}^-(\underline{z_n})\right]
\right\}_{\beta\alpha}\:
\nn\\
& \times\:
 \int \left[\prod_{n=1}^{N-1}  \frac{d^2 \p_n}{(2\pi)^2}\: e^{i \p_n\cdot (\z_{n+1}-\z_{n})}\,
 e^{-i\check{p}_n^-(z_{n+1}^+-z_{n}^+)}
 \right]
\, ,
\label{q_prop_pure_A_minus_Fourier_2}
\end{align}
where $p_n^+=k^+$ for $n$ from $1$ to $N\!-\!1$.

\subsubsection{Eikonal approximation for the quark propagator in a background gluon field}

The eikonal approximation can be performed at the level of the expression given in Eq. \eqref{q_prop_pure_A_minus_Fourier_2} as follows. Remembering that the eikonal approximation amounts, among other things, to the limit $L^+\rightarrow 0$ where $[-L^+/2,L^+/2]$ is the support of the background field $\mathcal{A}_a^-(\underline{z_n})$ along $z_n^+$, one has $|z_{n+1}^+\!-\!z_{n}^+|<L^+\rightarrow 0$, so that so phase factors $\exp{[-i\check{p}_n^-(z_{n+1}^+\!-\!z_{n}^+)]}$ can be replaced by $1$ in first approximation\footnote{Note that such an approximation can be performed only after taking the integrations in ${p}_n^-$. Otherwise, one would not be able to recover the correct ordering in $z^+$ unambigously.}. The integrals in $\p_n$ then reduce to $\delta^{(2)}(\z_{n+1}-\z_{n})$ factors, which leads to the expression
\begin{align}
\delta \tilde{S}_F(q,k)_{\beta\alpha}& \bigg|_{\textrm{pure }\mathcal{A}^{-}\textrm{, Eik.}}
= \, 2\pi\delta(q^+\!-\!k^+)\, \frac{i(\slashed{\check{q}}\!+\!m)}{[q^2\!-\!m^2+i\epsilon]}\gamma^+\frac{i(\slashed{\check{k}}\!+\!m)}{[k^2\!-\!m^2+i\epsilon]}
\int d^2\z\, e^{-i \z\cdot (\q-\k)}\,
\nn\\
& \times\:
\sum_{N=1}^{+\infty}\int\left[\prod_{n=1}^{N}  dz_n^+\right]\;
e^{i z_N^+ q^-}\, e^{-i z_1^+ k^-}\:
\left\{ \mathcal{P}_n \prod_{n=1}^{N} \left[-igt\!\cdot\!\mathcal{A}^-(z_n^+,\z)\right]
\right\}_{\beta\alpha}\:
\nn\\
& \times\:
\left\{ \theta(k^+) \left(\prod_{n=1}^{N-1}\theta(z_{n+1}^+\!-\!z_{n}^+)\right)
+(-1)^{N-1}\,\theta(-k^+) \left(\prod_{n=1}^{N-1}\theta(z_{n}^+\!-\!z_{n+1}^+)\right) \right\}\;
\label{q_prop_Eik_Fourier}
\end{align}
for the contribution of the interactions with $\mathcal{A}_a^-$ to the quark propagator, in the eikonal approximation. Interestingly, the $N=1$ term in Eq.~\eqref{q_prop_Eik_Fourier} has not been affected by the eikonal approximation, and is still exact in that sense.

Taking the Fourier transform of Eq.~\eqref{q_prop_Eik_Fourier}, following the convention in Eq. \eqref{q_prop_Fourier}, and performing the integrations over $q^-$ and $k^-$ thanks to Eq.~\eqref{p_minus_integration}, one finds
\begin{align}
\delta {S}_F&(x,y)_{\beta\alpha} \bigg|_{\textrm{pure }\mathcal{A}^{-}\textrm{, Eik.}}
=
\int \frac{d^3\uq}{(2\pi)^3} \int \frac{d^3\uk}{(2\pi)^3}\, 2\pi \delta(q^+\!-\!k^+)\,
e^{-i x\cdot \check{q}+i y\cdot \check{k}}\,
\frac{(\slashed{\check{q}}+m)\gamma^+(\slashed{\check{k}}+m)}{(2k^+)^2}\,
\nn\\
& \times\:
\int d^2\z\, e^{-i \z\cdot (\q-\k)}\,
\sum_{N=1}^{+\infty}\int\left[\prod_{n=1}^{N}  dz_n^+\right]\;
e^{i z_N^+ \check{q}^-}\, e^{-i z_1^+ \check{k}^-}\:
\left\{ \mathcal{P}_n \prod_{n=1}^{N} \left[-igt\!\cdot\!\mathcal{A}^-(z_n^+,\z)\right]
\right\}_{\beta\alpha}\:
\nn\\
& \times\:
\left\{ \theta(k^+) \left(\prod_{n=0}^{N}\theta(z_{n+1}^+\!-\!z_{n}^+)\right)
+(-1)^{N+1}\,\theta(-k^+) \left(\prod_{n=0}^{N}\theta(z_{n}^+\!-\!z_{n+1}^+)\right) \right\}
\, ,
\label{q_prop_Eik_generic_position_1}
\end{align}
where $z_{0}^+\equiv y^+$ and $z_{N+1}^+\equiv x^+$. Now that the integrations over $q^-$ and $k^-$ have been done, it is safe to neglect the phase factors dependent on $z_1^+$ and $z_N^+$, in the eikonal approximation. One can then recognize in Eq.~\eqref{q_prop_Eik_generic_position_1} the series expansion of the gauge link
\begin{align}
\mathcal{U}_F(x^+,y^+;\z)
\equiv &
\mathbf{1}
+\sum_{N=1}^{+\infty} \frac{1}{N!}\; \mathcal{P}_+\left[-ig\int_{y^+}^{x^+} dz^+\, t\!\cdot\! \mathcal{A}^-(z^+,\z) \right]^N
\nn\\
= &
\mathbf{1}
+\sum_{N=1}^{+\infty}\int\left[\prod_{n=1}^{N}  dz_n^+\right]\; \left(\prod_{n=0}^{N}\theta(z_{n+1}^+\!-\!z_{n}^+)\right)\;
\left\{ \mathcal{P}_n \prod_{n=1}^{N} \left[-igt\!\cdot\!\mathcal{A}^-(z_n^+,\z)\right]
\right\}
\label{Wilson_line}
\end{align}
and of the conjugate gauge link
\begin{align}
\mathcal{U}_F^{\dagger}(y^+,x^+;\z)
\equiv &
\mathbf{1}
+\sum_{N=1}^{+\infty} \frac{1}{N!}\; \left\{\overline{\mathcal{P}_+}\left[-ig\int_{x^+}^{y^+} dz^+\, t\!\cdot\! \mathcal{A}^-(z^+,\z) \right]^N\right\}^{\dagger}
\nn\\
= &
\mathbf{1}
+\sum_{N=1}^{+\infty}\int\left[\prod_{n=1}^{N}  dz_n^+\right]\; \left(\prod_{n=0}^{N}\theta(z_{n}^+\!-\!z_{n+1}^+)\right)\;
\left\{ \mathcal{P}_n \prod_{n=1}^{N} \left[+igt\!\cdot\!\mathcal{A}^-(z_n^+,\z)\right]
\right\}
\, ,
\label{Wilson_line_dagger}
\end{align}
where $\mathcal{P}_+$ and $\overline{\mathcal{P}_+}$ indicate respectively ordering and anti-ordering of the color matrices along $z^+$, from the right to the left.
In the eikonal approximation, the complete $\mathcal{A}^{-}$ medium correction to the quark propagator thus writes
\begin{align}
\delta {S}_F(x,y)_{\beta\alpha} \bigg|_{\textrm{pure }\mathcal{A}^{-}\textrm{, Eik.}}
=&
\int \frac{d^3\uq}{(2\pi)^3} \int \frac{d^3\uk}{(2\pi)^3}\, 2\pi \delta(q^+\!-\!k^+)\,
e^{-i x\cdot \check{q}+i y\cdot \check{k}}\,
\frac{(\slashed{\check{q}}+m)\gamma^+(\slashed{\check{k}}+m)}{(2k^+)^2}\,
\nn\\
 \times &\:
\int d^2\z\, e^{-i \z\cdot (\q-\k)}\,
\bigg\{ \theta(k^+) \theta(x^+\!-\!y^+)\, \Big[\mathcal{U}_F(x^+,y^+;\z)-\mathbf{1}\Big]_{\beta\alpha}
\nn\\
& \hspace{2cm}
- \theta(-k^+) \theta(y^+\!-\!x^+)\, \Big[\mathcal{U}_F^{\dagger}(y^+,x^+;\z)-\mathbf{1}\Big]_{\beta\alpha}
\bigg\}
\, .
\label{q_prop_Eik_generic_position_result}
\end{align}

The expression \eqref{q_prop_Eik_generic_position_result} is valid for all values of $x^+$ and $y^+$. Remember however that the background field $\mathcal{A}^-_a(\uz)$ is assumed to have a finite support $[-L^+/2, L^+/2]$ in $z^+$. The medium correction \eqref{q_prop_Eik_generic_position_result} receives a non-trivial contribution only from the overlap between the support $[-L^+/2, L^+/2]$ and the interval $[y^+, x^+]$ or $[x^+, y^+]$. In particular, when $x^+$ and $y^+$ are both before or both after the medium, the medium correction \eqref{q_prop_Eik_generic_position_result} vanishes, and the quark propagator reduces to its vacuum expression \eqref{free_q_prop_Fourier}, as expected for a Feynman propagator.

In general, the medium correction \eqref{q_prop_Eik_generic_position_result} can be combined with the vacuum propagator \eqref{free_q_prop_Fourier} in the following way. In Eq.~\eqref{free_q_prop_Fourier}, the momentum space propagator can be split as
\begin{align}
\frac{i(\slashed{k}+m)}{[k^2\!-\!m^2+i\epsilon]}
&= \frac{i\left[(\slashed{\check{k}}+m)+(k^-\!-\!\check{k}^-)\gamma^+\right]}{[k^2\!-\!m^2+i\epsilon]}
 = \frac{i(\slashed{\check{k}}+m)}{[k^2\!-\!m^2+i\epsilon]} + \frac{i\, \gamma^+}{2k^+}
\label{split_free_q_prop_pole_nonpole}
\end{align}
into a simple pole contribution in $k^-$ and a non-pole contribution. The simple pole contribution cancels identically the color identity matrix terms from Eq.~\eqref{q_prop_Eik_generic_position_result} and then, using the identity
\begin{align}
\int \frac{d k^+}{2\pi}\;  \frac{i}{2k^+}\, e^{-i k^+(x^-\!-\!y^-)}\: =&\: \frac{1}{4}\, \textrm{sgn}(x^-\!-\!y^-)
\label{int_sign_funct}
\end{align}
with the sign function $\textrm{sgn}(x)\equiv \theta(x)-\theta(-x)$, one obtains the expression
\begin{align}
{S}_F(x,y)_{\beta\alpha} \bigg|_{\textrm{pure }\mathcal{A}^{-}\textrm{, Eik.}}
=\, &
\mathbf{1}_{\alpha\beta}\; \delta^{(3)}(\ux\!-\!\uy)\; \textrm{sgn}(x^-\!-\!y^-)\; \frac{\gamma^+}{4}
\nn\\
  & +
\int \frac{d^3\uq}{(2\pi)^3} \int \frac{d^3\uk}{(2\pi)^3}\, 2\pi \delta(q^+\!-\!k^+)\,
e^{-i x\cdot \check{q}+i y\cdot \check{k}}\,
\frac{(\slashed{\check{q}}+m)\gamma^+(\slashed{\check{k}}+m)}{(2k^+)^2}\,
\nn\\
& \: \times \:
\int d^2\z\, e^{-i \z\cdot (\q-\k)}\,
\bigg\{ \theta(k^+) \theta(x^+\!-\!y^+)\, \mathcal{U}_F(x^+,y^+;\z)_{\beta\alpha}
\nn\\
& \hspace{3cm}
- \theta(-k^+) \theta(y^+\!-\!x^+)\, \mathcal{U}_F^{\dagger}(y^+,x^+;\z)_{\beta\alpha}
\bigg\}
\, .
\label{full_q_prop_Eik_generic_position_result}
\end{align}
for the quark propagator in an $\mathcal{A}^{-}$ background, valid in the eikonal approximation for any $x^+$ and $y^+$.

In specific regimes, Eq.~\eqref{full_q_prop_Eik_generic_position_result} may be approximated further. If, for example, $x^+$ belongs to the support $[-L^+/2, L^+/2]$, one can further enforce the eikonal approximation, by neglecting the $x^+$-dependent phase factors in Eq.~\eqref{full_q_prop_Eik_generic_position_result}, which leads to a $\delta^{(2)}(\x-\z)$ factor from the integration over $\q$. The situation is analog if $y^+$ belongs to $[-L^+/2, L^+/2]$.

\subsubsection{NEik quark propagator in a pure $\mathcal{A}^-$ background}

In order to go beyond the eikonal approximation, we follow a general strategy similar to the one used in Refs.~\cite{Altinoluk:2014oxa,Altinoluk:2015gia}, but with a few significant improvements and changes. In particular, we avoid the introduction of a discretization of the medium, and we deal with the incoming and outgoing quark in a more symmetric way.
The detailed derivation, starting from the general expression \eqref{q_prop_pure_A_minus_Fourier_2}, is exposed in Appendix \ref{appendix}. The obtained results for the cases of propagation through the whole medium (with pure $\mathcal{A}^{-}$ background field) are as follows.

For the quark propagator through the medium, corresponding to the case $x^+> L^+/2$ and $y^+<-L^+/2$, we obtain
\begin{align}
 S_F&(x,y) \bigg|_{\textrm{pure }\mathcal{A}^{-}}
=
\int \frac{d^3\uq}{(2\pi)^3} \int \frac{d^3\uk}{(2\pi)^3}\, 2\pi \delta(q^+\!-\!k^+)\, \frac{\theta(k^+)}{(2k^+)^2}\,
e^{-i x\cdot \check{q}+i y\cdot \check{k}}\,(\slashed{\check{q}}+m)\gamma^+(\slashed{\check{k}}+m)
\nn\\
\times &
\, \int d^2 \z\, e^{-i\z\cdot(\q\!-\!\k)} \,
\Bigg\{\mathcal{U}_F\Big(\frac{L^+}{2},-\frac{L^+}{2};\z\Big)
\nn\\
&
-\frac{(\q^j\!+\!\k^j)}{4k^+}\int_{-\frac{L^+}{2}}^{\frac{L^+}{2}}dz^+\, \left[\mathcal{U}_F\Big(\frac{L^+}{2},z^+;\z\Big)\,
\overleftrightarrow{\d_{\z^j}}\,
\mathcal{U}_F\Big(z^+,-\frac{L^+}{2};\z\Big)\right]
\nn\\
&
-\frac{i}{2k^+}\int_{-\frac{L^+}{2}}^{\frac{L^+}{2}}dz^+\,  \left[\mathcal{U}_F\Big(\frac{L^+}{2},z^+;\z\Big)\,
\overleftarrow{\d_{\z^j}}\, \overrightarrow{\d_{\z^j}}\,
\mathcal{U}_F\Big(z^+,-\frac{L^+}{2};\z\Big)\right]
\Bigg\}
+\textrm{NNEik}
\label{q_prop_ba_pure_A_minus}
\, .
\end{align}

By contrast, for the antiquark propagator through the medium, corresponding to the case $x^+< -L^+/2$ and $y^+>L^+/2$, we get
\begin{align}
 S_F&(x,y) \bigg|_{\textrm{pure }\mathcal{A}^{-}}
=
\int \frac{d^3\uq}{(2\pi)^3} \int \frac{d^3\uk}{(2\pi)^3}\, 2\pi \delta(q^+\!-\!k^+)\, \frac{\theta(-k^+)}{(2k^+)^2}\,
e^{-i x\cdot \check{q}+i y\cdot \check{k}}\,(\slashed{\check{q}}+m)\gamma^+(\slashed{\check{k}}+m)
\nn\\
\times &
\, \int d^2 \z\, e^{-i\z\cdot(\q\!-\!\k)} \,
\Bigg\{ - \mathcal{U}_F^{\dag}\Big(\frac{L^+}{2},-\frac{L^+}{2};\z\Big)
\nn\\
&
-\frac{(\q^j\!+\!\k^j)}{4k^+}\int_{-\frac{L^+}{2}}^{\frac{L^+}{2}}dz^+\, \left[\mathcal{U}_F^{\dag}\Big(z^+,-\frac{L^+}{2};\z\Big)\,
\overleftrightarrow{\d_{\z^j}}\,
\mathcal{U}_F^{\dag}\Big(\frac{L^+}{2},z^+;\z\Big)
\right]
\nn\\
&
-\frac{i}{2k^+}\int_{-\frac{L^+}{2}}^{\frac{L^+}{2}}dz^+\,  \left[\mathcal{U}_F^{\dag}\Big(z^+,-\frac{L^+}{2};\z\Big)\,
\overleftarrow{\d_{\z^j}}\, \overrightarrow{\d_{\z^j}}\,
\mathcal{U}_F^{\dag}\Big(\frac{L^+}{2},z^+;\z\Big)\right]
\Bigg\}
+\textrm{NNEik}
\label{qbar_prop_ba_pure_A_minus}
\, .
\end{align}

Note that in both Eqs.~\eqref{q_prop_ba_pure_A_minus} and \eqref{qbar_prop_ba_pure_A_minus}, the transverse derivatives act only on the Wilson lines, not on the phase factor.

\subsection{Single $\mathcal{A}_{\perp}$ insertion contribution}

Once the quark propagator in a pure $\mathcal{A}^-$ background is known, the corrections to the quark propagator due to the interactions with other components of the background field can be formulated via perturbation theory in position space. For example, the contribution of a single insertion of the transverse components writes
\begin{align}
\delta S_F(x,y)\bigg|_{\textrm{single }\mathcal{A}_{\perp}}
& = \int d^4z\; S_F(x,z) \bigg|_{\textrm{pure }\mathcal{A}^{-}}\;
[-ig\, \gamma^j\, t^a]\; \mathcal{A}^{a}_{j}(\uz)\;\;\;
S_F(z,y) \bigg|_{\textrm{pure }\mathcal{A}^{-}}
\, .
\label{single_A_perp_term_def}
\end{align}
In Eq.~\eqref{single_A_perp_term_def}, the integration in $z^+$ amounts to a $L^+$ factor for $L^+\rightarrow 0$, whereas $\mathcal{A}^{a}_{j}$ is independent of $L^+$ in that limit. Hence, the expression \eqref{single_A_perp_term_def} seems to start at Next-to-Eikonal accuracy. However, such naive power counting fails due to the instantaneous contribution to the quark propagator, see Eq.~\eqref{full_q_prop_Eik_generic_position_result}, since delta distribution makes the integration in $z^+$ trivial. In that case, the transverse field $\mathcal{A}^{a}_{j}(\uz)$ is taken at $z^+=x^+$ or at $z^+=y^+$, which is possible only if $x^+$ or $y^+$ belong to the support $[-L^+/2, L^+/2]$.

In order to avoid these complications, we will focus on the case in which neither $x^+$ nor $y^+$ belong to the support $[-L^+/2, L^+/2]$ of the background field, so that the naive power counting in $L^+$ applies. Then, Eq.~\eqref{single_A_perp_term_def} starts at NEik accuracy, with a term obtained by using the eikonal expression
\eqref{full_q_prop_Eik_generic_position_result} for both propagators in $\mathcal{A}^-$ background.

For example, for $x^+> L^+/2$ and $y^+< -L^+/2$, one obtains
\begin{align}
\delta S_F&(x,y)\bigg|_{\textrm{single }\mathcal{A}_{\perp}}
 = \int d^4z
\int \frac{d^3\uq}{(2\pi)^3} \int \frac{d^3\underline{p_2}}{(2\pi)^3}\, 2\pi \delta(q^+\!-\!p_2^+)\,
e^{-i x\cdot \check{q}+i z\cdot \check{p}_2}\,
\frac{(\slashed{\check{q}}+m)\gamma^+(\slashed{\check{p}}_2+m)}{(2q^+)^2}
\nn\\
& \times \:
\int d^2\v\, e^{-i \v\cdot (\q-\p_2)}\,
\theta(q^+) \, \mathcal{U}_F\left(\frac{L^+}{2},z^+;\v\right)\:
[-ig\, \gamma^j\, t^a]\; \mathcal{A}^{a}_{j}(\uz)\;\;\;
\nn\\
& \times \:
\int \frac{d^3\underline{p_1}}{(2\pi)^3} \int \frac{d^3\uk}{(2\pi)^3}\, 2\pi \delta(p_1^+\!-\!k^+)\,
e^{-i z\cdot \check{p}_1+i y\cdot \check{k}}\,
\frac{(\slashed{\check{p}}_1+m)\gamma^+(\slashed{\check{k}}+m)}{(2k^+)^2}\,
\nn\\
& \times \:
\int d^2\u\, e^{-i \u\cdot (\p_1-\k)}\,
 \theta(k^+) \, \mathcal{U}_F\left(z^+,-\frac{L^+}{2};\u\right)\;\;\;
+\;\textrm{NNEik}
\, ,
\label{single_A_perp_term_1}
\end{align}
since by hypothesis $x^+>z^+>y^+$, so that only the terms with positive light-cone momentum $q^+$ or $k^+$ from Eq.~\eqref{full_q_prop_Eik_generic_position_result} survive.
Then, the integration over $z^-$ forces $p_1^+=p_2^+=k^+=q^+$. The Dirac algebra in Eq.~\eqref{single_A_perp_term_1} can then be performed as
\begin{align}
(\slashed{\check{q}}+m)\gamma^+(\slashed{\check{p}}_2+m)\gamma^j(\slashed{\check{p}}_1+m)\gamma^+(\slashed{\check{k}}+m)
=&
(2k^+)\, (\slashed{\check{q}}+m)\left[\gamma^j\gamma^+\gamma^i\,\p_1^i + \p_2^i\,\gamma^i\gamma^+\gamma^j\right](\slashed{\check{k}}+m)
\label{Dirac_num_single_A_perp_insertion}
\, .
\end{align}
Representing $\p_1^i$ and $\p_2^i$ in the numerator \eqref{Dirac_num_single_A_perp_insertion} in terms of derivatives with respect to $\u^i$ or $\v^i$ acting on the phase factor in Eq.~\eqref{single_A_perp_term_1}, one obtains after elementary calculations the expression
\begin{align}
\delta S_F&(x,y) \bigg|_{\textrm{single }\mathcal{A}_{\perp}}
=
\int \frac{d^3\uq}{(2\pi)^3} \int \frac{d^3\uk}{(2\pi)^3}\, 2\pi \delta(q^+\!-\!k^+)\, \frac{\theta(k^+)}{(2k^+)^3}\,
e^{-i x\cdot \check{q}}\,e^{i y\cdot \check{k}}
\nn\\
& \times
\, (\slashed{\check{q}}+m)\gamma^j\gamma^+\gamma^i\,(\slashed{\check{k}}+m) \,
\int d^3 \uz\,  \left[e^{-i\z\cdot\q}\;\mathcal{U}_F\Big(\frac{L^+}{2},z^+;\z\Big)\right]
\nn\\
& \times\;
\bigg[
 \overleftarrow{\d_{\z^j}}
\big[gt\!\cdot\!\mathcal{A}_{i}(\uz)\big]
-
\big[gt\!\cdot\!\mathcal{A}_{j}(\uz)\big]
\overrightarrow{\d_{\z^i}}
\bigg]
\left[\mathcal{U}_F\Big(z^+,-\frac{L^+}{2};\z\Big)\;
e^{i\z\cdot\k}\right]\;\;\;
+\;\textrm{NNEik}
\label{q_prop_ba_single_A_perp}
\end{align}
for the NEik correction to the quark propagator through a shockwave due to single interaction with the transverse component of the background field, for $x^+> L^+/2$ and $y^+< -L^+/2$ as a reminder.
Note that in Eq.~\eqref{q_prop_ba_single_A_perp}, the transverse derivatives act both on the Wilson lines and on the phase factors.

The case of antiquark propagating through a shockwave, corresponding to $y^+> L^+/2$ and $x^+< -L^+/2$ instead, can be calculated in a similar way, but keeping terms with negative $q^+$ or $k^+$ from Eq.~\eqref{full_q_prop_Eik_generic_position_result} this time. One obtain this time
\begin{align}
\delta S_F&(x,y) \bigg|_{\textrm{single }\mathcal{A}_{\perp}}
=
\int \frac{d^3\uq}{(2\pi)^3} \int \frac{d^3\uk}{(2\pi)^3}\, 2\pi \delta(q^+\!-\!k^+)\, \frac{\theta(-k^+)}{(2k^+)^3}\,
e^{-i x\cdot \check{q}}\,e^{i y\cdot \check{k}}
\nn\\
& \times
\, (\slashed{\check{q}}+m)\gamma^j\gamma^+\gamma^i\,(\slashed{\check{k}}+m) \,
\int d^3 \uz\,  \left[e^{-i\z\cdot\q}\;\mathcal{U}_F^{\dag}\Big(z^+,-\frac{L^+}{2};\z\Big)\right]
\nn\\
& \times\;
\bigg[
 \overleftarrow{\d_{\z^j}}
\big[gt\!\cdot\!\mathcal{A}_{i}(\uz)\big]
-
\big[gt\!\cdot\!\mathcal{A}_{j}(\uz)\big]
\overrightarrow{\d_{\z^i}}
\bigg]
\left[\mathcal{U}_F^{\dag}\Big(\frac{L^+}{2},z^+;\z\Big)\;
e^{i\z\cdot\k}\right]\;\;\;
+\;\textrm{NNEik}
\, .
\label{qbar_prop_ba_single_A_perp}
\end{align}

\subsection{Instantaneous double $\mathcal{A}_{\perp}$ insertion contribution}

One can go further and consider the contributions to the propagator with two or more insertions of the transverse components of the background field. For example, for two insertions, one has
\begin{align}
\delta S_F(x,y)\bigg|_{\textrm{double }\mathcal{A}_{\perp}}
=
 \int d^4z\;\int d^4z'&\; S_F(x,z') \bigg|_{\textrm{pure }\mathcal{A}^{-}}\;
[-ig\, \gamma^j\, t^b]\; \mathcal{A}^{b}_{j}(\uz')\;\;\;
S_F(z',z) \bigg|_{\textrm{pure }\mathcal{A}^{-}}
\nn\\
& \times \;
[-ig\, \gamma^i\, t^a]\; \mathcal{A}^{a}_{i}(\uz)\;\;\;
S_F(z,y) \bigg|_{\textrm{pure }\mathcal{A}^{-}}
\, .
\label{double_A_perp_term_def}
\end{align}
By naive power counting, each extra insertion brings an extra power of $L^+$ for $L^+\rightarrow 0$, due to the extra integration. As we discussed in the previous section, this naive power counting fails due to the instantaneous contribution to the quark propagator, see Eq.~\eqref{full_q_prop_Eik_generic_position_result}. Again, we focus on the cases in which neither $x^+$ nor $y^+$ belong to the support $[-L^+/2, L^+/2]$ of the background field, so that the instantaneous term in Eq.~\eqref{full_q_prop_Eik_generic_position_result} does not contribute to the external legs in Eq.~\eqref{double_A_perp_term_def}. By contrast, the instantaneous term contributes to the internal line, giving the dominant contribution to Eq.~\eqref{double_A_perp_term_def}, which is at  NEik accuracy.

In particular, for $x^+> L^+/2$ and $y^+< -L^+/2$, one has
\begin{align}
\delta S_F&(x,y)\bigg|_{\textrm{double }\mathcal{A}_{\perp}}
 =  \int d^4z\;\int d^4z'\;
\int \frac{d^3\uq}{(2\pi)^3} \int \frac{d^3\underline{p_2}}{(2\pi)^3}\, 2\pi \delta(q^+\!-\!p_2^+)\,
e^{-i x\cdot \check{q}+i z'\cdot \check{p}_2}\,
\nn\\
& \times \: \frac{(\slashed{\check{q}}+m)\gamma^+(\slashed{\check{p}}_2+m)}{(2q^+)^2}
\int d^2\v\, e^{-i \v\cdot (\q-\p_2)}\,
\theta(q^+) \, \mathcal{U}_F\left(\frac{L^+}{2},{z'}^+;\v\right)\:
\nn\\
& \times \: [-ig\, \gamma^j\, t^b]\; \mathcal{A}^{b}_{j}(\uz')\;\;\;
\delta^{(3)}(\uz'\!-\!\uz)\, \gamma^+\, \int \frac{d p^+}{2\pi}\;  \frac{i}{2p^+}\, e^{-i p^+({z'}^-\!-\!z^-)}\:
[-ig\, \gamma^i\, t^a]\; \mathcal{A}^{a}_{i}(\uz)\;\;\;
\nn\\
& \times \:
\int \frac{d^3\underline{p_1}}{(2\pi)^3} \int \frac{d^3\uk}{(2\pi)^3}\, 2\pi \delta(p_1^+\!-\!k^+)\,
e^{-i z\cdot \check{p}_1+i y\cdot \check{k}}\,
\frac{(\slashed{\check{p}}_1+m)\gamma^+(\slashed{\check{k}}+m)}{(2k^+)^2}\,
\nn\\
& \times \:
\int d^2\u\, e^{-i \u\cdot (\p_1-\k)}\,
 \theta(k^+) \, \mathcal{U}_F\left(z^+,-\frac{L^+}{2};\u\right)\;\;\;
+\;\textrm{NNEik}
\, ,
\label{double_A_perp_term_1}
\end{align}
using the integral representation \eqref{int_sign_funct} for the sign function. It is then straightforward to obtain
\begin{align}
\delta S_F(x,y) &\bigg|_{\textrm{double }\mathcal{A}_{\perp}}
=
\int \frac{d^3\uq}{(2\pi)^3} \int \frac{d^3\uk}{(2\pi)^3}\, 2\pi \delta(q^+\!-\!k^+)\, \frac{\theta(k^+)}{(2k^+)^3}\,
e^{-i x\cdot \check{q}}\, e^{i y\cdot \check{k}}
\nn\\
& \times
\, (\slashed{\check{q}}+m)\gamma^j\gamma^+\gamma^i\,(\slashed{\check{k}}+m) \,
\int d^3 \uz\, e^{-i\z\cdot(\q\!-\!\k)}
\nn\\
& \times\;
(-i)\;\;
\mathcal{U}_F\Big(\frac{L^+}{2},z^+;\z\Big)\,
\big[gt\!\cdot\!\mathcal{A}_{j}(\uz)\big]\big[gt\!\cdot\!\mathcal{A}_{i}(\uz)\big]
\mathcal{U}_F\Big(z^+,-\frac{L^+}{2};\z\Big)
+\textrm{NNEik}
\label{q_prop_ba_double_A_perp}
\end{align}
for $x^+> L^+/2$ and $y^+< -L^+/2$, corresponding to the case of a quark going through the whole shockwave.

Similarly, for $y^+> L^+/2$ and $x^+< -L^+/2$, corresponding to the case of an antiquark going through the whole shockwave  one finds
\begin{align}
\delta S_F(x,y) &\bigg|_{\textrm{double }\mathcal{A}_{\perp}}
=
\int \frac{d^3\uq}{(2\pi)^3} \int \frac{d^3\uk}{(2\pi)^3}\, 2\pi \delta(q^+\!-\!k^+)\, \frac{\theta(-k^+)}{(2k^+)^3}\,
e^{-i x\cdot \check{q}}\, e^{i y\cdot \check{k}}
\nn\\
& \times
\, (\slashed{\check{q}}+m)\gamma^j\gamma^+\gamma^i\,(\slashed{\check{k}}+m) \,
\int d^3 \uz\, e^{-i\z\cdot(\q\!-\!\k)}
\nn\\
& \times\;
(-i)\;\;
\mathcal{U}_F^{\dag}\Big(z^+,-\frac{L^+}{2};\z\Big)\,
\big[gt\!\cdot\!\mathcal{A}_{j}(\uz)\big]\big[gt\!\cdot\!\mathcal{A}_{i}(\uz)\big]
\mathcal{U}_F^{\dag}\Big(\frac{L^+}{2},z^+;\z\Big)
+\textrm{NNEik}
\, .
\label{qbar_prop_ba_double_A_perp}
\end{align}

A priori, one could generalize such results to the case of higher number of insertions of the transverse components of the background field. Keeping only the instantaneous term in the internal quark lines, only one integration in $z^+$
would survive, thus giving a Next-to-Eikonal contribution. However, that contribution vanishes, because the instantaneous term is proportional to $\gamma^+$, which anticommutes with the transverse $\gamma^j$. Hence, there is no non-vanishing contribution with two successive instantaneous quark lines separated only by a $\mathcal{A}_{\perp}$ insertion.

\subsection{Full result for the NEik quark or antiquark propagator through a shockwave}

So far, we have obtained the expression \eqref{q_prop_ba_pure_A_minus} for the quark propagator through a pure $\mathcal{A}^-$ background field at NEik accuracy, as well as the NEik corrections due to single or double $\mathcal{A}_{\perp}$ contributions, given in Eqs.~\eqref{q_prop_ba_single_A_perp} and \eqref{q_prop_ba_double_A_perp} respectively. The pure $\mathcal{A}^-$ contributions involve the same spinor structure $(\slashed{\check{q}}+m)\gamma^+(\slashed{\check{k}}+m)$ at Eikonal and NEik accuracy. By contrast, the single and double $\mathcal{A}_{\perp}$ contributions involve instead the spinor structure $(\slashed{\check{q}}+m)\gamma^j\gamma^+\gamma^i\,(\slashed{\check{k}}+m)$. It turns out convenient to separate the symmetric and antisymmetric terms in $i,j$ in that spinor structure, as
\begin{align}
\gamma^j\gamma^+\gamma^i=-\gamma^+\gamma^j\gamma^i
= &
-\gamma^+\left(\frac{\{\gamma^j,\gamma^i\}}{2}+\frac{[\gamma^j,\gamma^i]}{2}\right)
=\delta^{ij}\, \gamma^+ + \gamma^+\frac{[\gamma^i,\gamma^j]}{2}
\, ,
\end{align}
so that the symmetric part can be combined with the pure $\mathcal{A}^-$ contributions. At NEik accuracy, the propagator can thus be split as
\begin{align}
S_F(x,y) = & S_F(x,y) \bigg|_{\textrm{unpol.}} + S_F(x,y) \bigg|_{\textrm{h. dep.}}
\label{unpol_hdep_decomp}
\end{align}
into an unpolarized piece and an helicity-dependent piece, since the $[\gamma^i,\gamma^j]$ is proportional to the quark helicity when acting on a $u$ or $v$ spinor.

Collecting and combining the contributions from Eqs.~\eqref{q_prop_ba_pure_A_minus}, \eqref{q_prop_ba_single_A_perp} and \eqref{q_prop_ba_double_A_perp}. We find for the quark propagator through the shock, for $x^+> L^+/2$ and $y^+< -L^+/2$,
\begin{align}
 S_F&(x,y) \bigg|_{\textrm{unpol.}}
=
\int \frac{d^3\uq}{(2\pi)^3} \int \frac{d^3\uk}{(2\pi)^3}\, 2\pi \delta(q^+\!-\!k^+)\, \frac{\theta(k^+)}{(2k^+)^2}\,
e^{-i x\cdot \check{q}}\, e^{i y\cdot \check{k}}\,(\slashed{\check{q}}+m)\gamma^+(\slashed{\check{k}}+m)
\nn\\
\times &
\, \int d^2 \z\, e^{-i\z\cdot(\q\!-\!\k)} \,
\Bigg\{\mathcal{U}_F\Big(\frac{L^+}{2},-\frac{L^+}{2};\z\Big)
\nn\\
&
-\frac{(\q^j\!+\!\k^j)}{4k^+}\int_{-\frac{L^+}{2}}^{\frac{L^+}{2}}dz^+\, \left[\mathcal{U}_F\Big(\frac{L^+}{2},z^+;\z\Big)\,
\overleftrightarrow{\mathcal{D}_{\z^j}}\,
\mathcal{U}_F\Big(z^+,-\frac{L^+}{2};\z\Big)\right]
\nn\\
&
-\frac{i}{2k^+}\int_{-\frac{L^+}{2}}^{\frac{L^+}{2}}dz^+\,  \left[\mathcal{U}_F\Big(\frac{L^+}{2},z^+;\z\Big)\,
\overleftarrow{\mathcal{D}_{\z^j}}\, \overrightarrow{\mathcal{D}_{\z^j}}\,
\mathcal{U}_F\Big(z^+,-\frac{L^+}{2};\z\Big)\right]
\Bigg\}
+\textrm{NNEik}
\label{q_prop_ba_h_indep}
\end{align}
and
\begin{align}
& S_F(x,y) \bigg|_{\textrm{h. dep.}}
=
\int \frac{d^3\uq}{(2\pi)^3} \int \frac{d^3\uk}{(2\pi)^3}\, 2\pi \delta(q^+\!-\!k^+)\, \frac{\theta(k^+)}{(2k^+)^3}\,
e^{-i x\cdot \check{q}}\, e^{i y\cdot \check{k}}
\nn\\
& \times
\, (\slashed{\check{q}}+m)\gamma^+ \frac{[\gamma^i,\gamma^j]}{4}\,(\slashed{\check{k}}+m) \,
\int d^2 \z\, e^{-i\z\cdot(\q\!-\!\k)}
\;
\nn\\
& \times
\int_{-\frac{L^+}{2}}^{\frac{L^+}{2}}dz^+\,
\mathcal{U}_F\Big(\frac{L^+}{2},z^+;\z\Big)\,
g t\!\cdot\!\mathcal{F}_{ij}(\uz)\,
\mathcal{U}_F\Big(z^+,-\frac{L^+}{2};\z\Big)+\textrm{NNEik}
\label{q_prop_ba_h_dep}
\, ,
\end{align}

with the notations
\begin{align}
\overrightarrow{\mathcal{D}_{z^{\mu}}} \equiv &\, \overrightarrow{\d_{z^{\mu}}} +igt\!\cdot\!\mathcal{A}_{\mu}(\uz) \\
\overleftarrow{\mathcal{D}_{z^{\mu}}} \equiv &\, \overleftarrow{\d_{z^{\mu}}} -igt\!\cdot\!\mathcal{A}_{\mu}(\uz) \\
\overleftrightarrow{\mathcal{D}_{z^{\mu}}} \equiv &\, \overrightarrow{\mathcal{D}_{z^{\mu}}} - \overleftarrow{\mathcal{D}_{z^{\mu}}} = \overleftrightarrow{\d_{z^{\mu}}} +2igt\!\cdot\!\mathcal{A}_{\mu}(\uz)\\
\mathcal{F}^a_{ij}(\uz)\equiv &\, \d_{\z^i}\mathcal{A}^a_{j}(\uz)-\d_{\z^j}\mathcal{A}^a_{i}(\uz) -g f^{abc}\, \mathcal{A}^b_{i}(\uz)\, \mathcal{A}^c_{j}(\uz)
\end{align}
for the background covariant derivatives and field strength. In Eq.~\eqref{q_prop_ba_h_indep} the transverse covariant derivatives act only on the Wilson lines, not on the phase factors. The results \eqref{q_prop_ba_h_indep} and \eqref{q_prop_ba_h_dep} are written in terms of gauge-covariant building blocks as expected, since no explicit gauge condition has been imposed on the background field.

For the case of an antiquark crossing the shockwave, meaning $y^+> L^+/2$ and $x^+< -L^+/2$, the decomposition \eqref{unpol_hdep_decomp} is still valid, but the two contributions, obtained from Eqs.~\eqref{qbar_prop_ba_pure_A_minus}, \eqref{qbar_prop_ba_single_A_perp} and \eqref{qbar_prop_ba_double_A_perp}, are now
\begin{align}
 S_F&(x,y) \bigg|_{\textrm{unpol.}}
=
\int \frac{d^3\uq}{(2\pi)^3} \int \frac{d^3\uk}{(2\pi)^3}\, 2\pi \delta(q^+\!-\!k^+)\, \frac{\theta(-k^+)}{(2k^+)^2}\,
e^{-i x\cdot \check{q}}\, e^{i y\cdot \check{k}}\,(\slashed{\check{q}}+m)\gamma^+(\slashed{\check{k}}+m)
\nn\\
\times &
\, \int d^2 \z\, e^{-i\z\cdot(\q\!-\!\k)} \,
\Bigg\{-\mathcal{U}_F^{\dag}\Big(\frac{L^+}{2},-\frac{L^+}{2};\z\Big)
\nn\\
&
-\frac{(\q^j\!+\!\k^j)}{4k^+}\int_{-\frac{L^+}{2}}^{\frac{L^+}{2}}dz^+\,
\left[\mathcal{U}_F^{\dag}\Big(z^+,-\frac{L^+}{2};\z\Big)\,
\overleftrightarrow{\mathcal{D}_{\z^j}}\,
\mathcal{U}_F^{\dag}\Big(\frac{L^+}{2},z^+;\z\Big)\right]
\nn\\
&
-\frac{i}{2k^+}\int_{-\frac{L^+}{2}}^{\frac{L^+}{2}}dz^+\,
\left[\mathcal{U}_F^{\dag}\Big(z^+,-\frac{L^+}{2};\z\Big)\,
\overleftarrow{\mathcal{D}_{\z^j}}\, \overrightarrow{\mathcal{D}_{\z^j}}\,
\mathcal{U}_F^{\dag}\Big(\frac{L^+}{2},z^+;\z\Big)\right]
\Bigg\}
+\textrm{NNEik}
\label{qbar_prop_ba_h_indep}
\end{align}

and
\begin{align}
& S_F(x,y) \bigg|_{\textrm{h. dep.}}
=
\int \frac{d^3\uq}{(2\pi)^3} \int \frac{d^3\uk}{(2\pi)^3}\, 2\pi \delta(q^+\!-\!k^+)\, \frac{\theta(-k^+)}{(2k^+)^3}\,
e^{-i x\cdot \check{q}}\, e^{i y\cdot \check{k}}
\nn\\
& \times
\, (\slashed{\check{q}}+m)\gamma^+ \frac{[\gamma^i,\gamma^j]}{4}\,(\slashed{\check{k}}+m) \,
\int d^2 \z\, e^{-i\z\cdot(\q\!-\!\k)}
\;
\nn\\
& \times
\int_{-\frac{L^+}{2}}^{\frac{L^+}{2}}dz^+\,
\mathcal{U}_F^{\dag}\Big(z^+,-\frac{L^+}{2};\z\Big)\,
g t\!\cdot\!\mathcal{F}_{ij}(\uz)\,
\mathcal{U}_F^{\dag}\Big(\frac{L^+}{2},z^+;\z\Big)+\textrm{NNEik}
\label{qbar_prop_ba_h_dep}
\, .
\end{align}

Before applying our results to the forward quark-nucleus scattering, a few comments are in order. The helicity dependent part of the  propagator in the gluon background given in Eqs. \eqref{q_prop_ba_h_dep} and \eqref{qbar_prop_ba_h_dep} for the quark and antiquark cases, is in agreement with the results derived and used in \cite{Kovchegov:2017lsr, Kovchegov:2018znm, Cougoulic:2019aja, Cougoulic:2020tbc} as well as in \cite{Chirilli:2018kkw, Chirilli:2021lif}. Concerning the unpolarized part of the propagator given in Eqs. \eqref{q_prop_ba_h_indep} and \eqref{qbar_prop_ba_h_indep} for the quark and antiquark cases, a full comparison between the results of the present manuscript and the ones from \cite{Chirilli:2018kkw} is rather complicated. This is due to the fact that different methods were used in the derivation, which in turn lead to results written in a different form. In Ref. \cite{Chirilli:2021lif}, the multiple differences in the results were claimed to be genuine disagreements. However, as argued in Appendix \ref{Sec:Comparison}, at least part of these apparent differences are irrelevant. On the other hand, a full comparison of the two different methods employed in \cite{Chirilli:2018kkw} and the present manuscript is essential in order to fully settle this issue. Due to its complexity, this task is beyond the scope the present paper. A dedicated study of this issue is currently in progress and will be presented in \cite{Altinoluk_Beuf}.  





\section{Forward quark-nucleus scattering at NEik accuracy}
\label{sec:cross_section}

This section is devoted to the application of our results derived in Section \ref{sec:quark_prop} to a specific observable, namely single inclusive forward quark production in pA collisions. One can argue that the forward quark production occurs at high energies and therefore NEik corrections, that are energy suppressed, would be negligible for this specific observable. However, at moderate energies these corrections might be vital for precise description of the experimental data especially from RHIC and the future EIC. In order to increase the precision, one should consider not only the NLO corrections in the strong coupling constant but also the NEik corrections when describing the experimental data for single inclusive production at moderate energies.
Moreover, the NEik corrections computed for forward quark production in this section are expected to provide a benchmark for the energy scale where one can safely use eikonal approximation and neglect the energy suppressed corrections. This will be realized by a numerical comparison of the NEik corrections to the strict eikonal expression (similar to the study performed in \cite{Agostini:2019avp}) which we leave as a future work.

In the rest of this section, first we compute the quark-nucleus scattering amplitude by using the LSZ reduction formula and then use the computed amplitude to calculate the unpolarized forward quark production cross section as well as the quark helicity asymmetry at NEik accuracy. The analogous study for antiquarks is presented in Appendix \ref{Sec:antiquark-observables}.

\subsection{Quark-target scattering amplitude from LSZ reduction}
\label{sec:amplitude}

In order to compute the quark-target scattering amplitude, we start from the definition of the free fermion fields which are given by
\ba
\label{Psi-1}
\Psi_\alpha (x) & =& \int_0^\infty \frac{dp^+}{2\pi} \int \frac{d^2 \p}{(2\pi)^2} \frac{1}{2p^+}
\sum_h \Big[ \hat b(\check{p},h,\alpha) u(\check{p},h) e^{-i\check{p}\cdot x}
+ \hat d^\dagger (\check{p},h,\alpha) v(\check{p},h) e^{i\check{p}\cdot x} \Big] \, \quad \quad \\
\label{Psi-2}
\bar \Psi_\alpha (x) & =& \int_{0}^\infty \frac{dp^+}{2\pi} \int \frac{d^2\p}{(2\pi)^2} \frac{1}{2p^+}
\sum_h \Big[ \hat d(\check{p},h,\alpha) \bar v(\check{p},h) e^{-i\check{p}\cdot x}
+ \hat b^\dagger (\check{p},h,\alpha) \bar u(\check{p},h) e^{i\check{p}\cdot x} \Big] \,
\ea
where $\check{p}$ is the on-shell momentum such that $\check{p}^-=(\p^2+m^2)/(2p^+)$ and $h=\pm 1/2$ is helicity. $\hat b$ and $\hat b^\dagger$ ($\hat d$ and $\hat d^\dagger$) are the annihilation and creation operators, respectively, of a quark (antiquark).
Using the property
\ba
\label{eq:condition-spinors}
\bar u(\check{p},h) \gamma^+ u(\check{k},h') = \sqrt{2p^+}\sqrt{2k^+} \delta_{hh'}
\ea
of the spinors, Eqs.~\eqref{Psi-1} and \eqref{Psi-2} can be inverted in order to express the annihilation and creation operators for a quark in terms of the free quark field as
\ba
\label{Creation_Gen}
\hat b(\check{k},h,\alpha) &=& \int d^2 \x \int dx^- e^{ix\cdot \check{k}}
\bar u(\check{k},h) \gamma^+ \Psi_\alpha(x), \\
\label{Annihilation_Gen}
\hat b^\dagger(\check{k},h,\alpha) &=& \int d^2 \x \int dx^- e^{-ix\cdot \check{k}}
\bar \Psi_\alpha(x) \gamma^+ u(\check{k},h).
\ea
%

The process we would like to study is a quark that undergoes multiple scatterings while propagating through a dense target that is defined in terms of strong background field.
As usual, within perturbation theory, the quark is considered free in the asymptotic past and asymptotic future, allowing to build the corresponding "${\rm in}$" and "${\rm out}$" Fock space.
In particular, the annihilation operator of a quark in the "${\rm out}$" Fock space and creation operator of a quark in the "${\rm in}$" Fock space can be written as
\ba
\label{eq:b-out}
\hat b_{\rm out}(\check{q},h',\beta) &=&  \lim_{x^+\to +\infty} \int d^2 \x \int dx^- e^{ix\cdot \check{q}}
\bar u(\check{q},h') \gamma^+ \Psi_\beta(x), \\
\label{eq:b-in}
\hat b_{\rm in}^\dagger(\check{k},h,\alpha) &=& \lim_{y^+\to -\infty} \int d^2 \y\int dy^- e^{-iy\cdot \check{k}}
\bar \Psi_\alpha(y) \gamma^+ u(\check{k},h)
\ea
now with the interacting quark field $\Psi_\beta(x)$.
%
%

\begin{figure}[!t]
\centering
\includegraphics[scale=0.7]{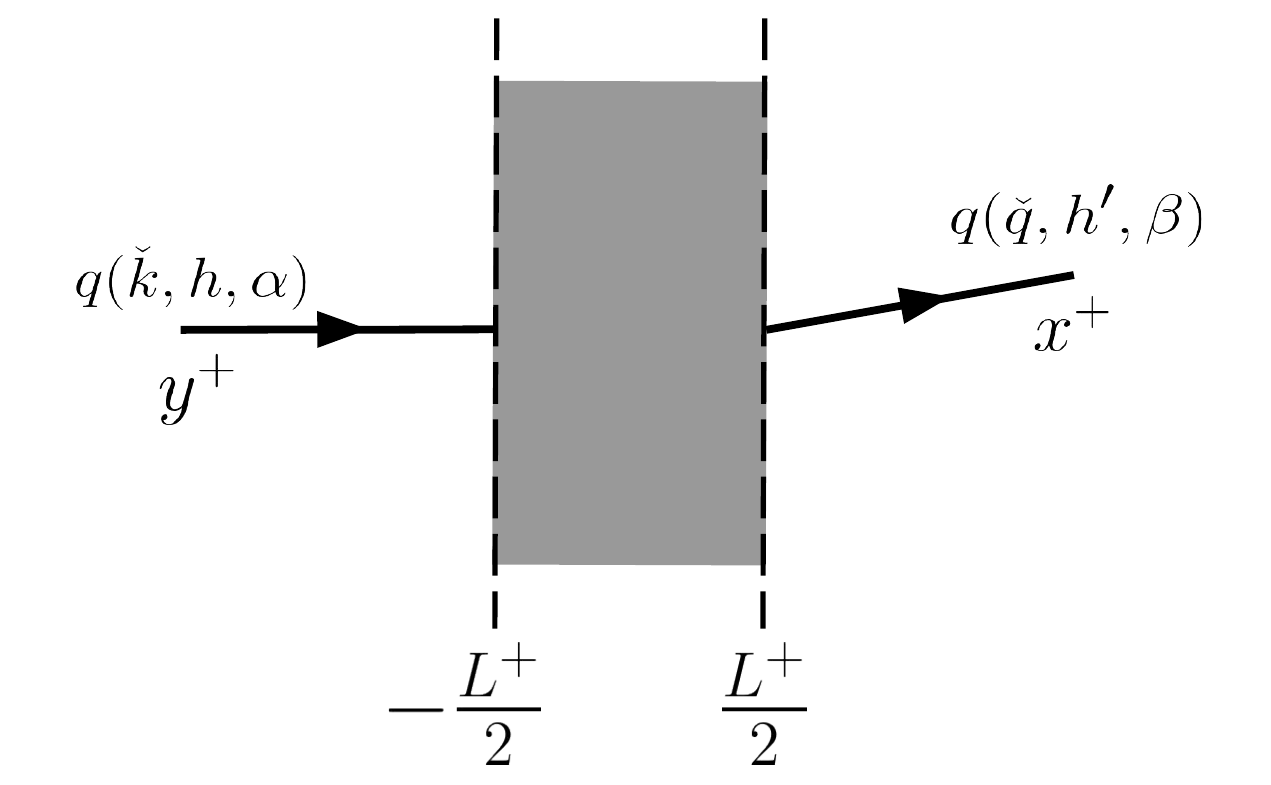}
\caption{Elastic scattering of a quark on a strong background field.}
\label{fig:quark}
\end{figure}

The $S$-matrix element for quark scattering on the background field (see Fig.~\ref{fig:quark}) is defined as
\ba
\label{eq:S-matrix-def}
S_{q(\check{q},h',\beta)\leftarrow q(\check{k},h,\alpha) }  &=& \langle 0| \hat b_{\rm out} (\check{q},h,\beta)
\hat b^\dagger_{\rm in} (\check{k}, h,\alpha) | 0 \rangle
\ea
where we denote the vacuum states by $\langle 0|=\langle 0|_{\rm in}=\langle 0|_{\rm out}$. By using the definitions of the incoming and outgoing quark creation and annihilation operators given in Eqs.~\eqref{eq:b-out} and \eqref{eq:b-in}, the $S$-matrix element can be cast into the following form:
\ba
\label{eq:S-matrix}
S_{q(\check{q},h',\beta)\leftarrow q(\check{k},h,\alpha) } &=& \lim_{x^+\to +\infty} \lim_{y^+ \to -\infty} \int d^2 \x \int dx^- \! \! \int d^2 \y \int dy^-
e^{ix\cdot {\check{q}} - iy\cdot {\check{k}} }  \nn \\
&& \qquad\qquad \times
\bar u(\check{q},h') \gamma^+ S_F(x,y)_{\beta\alpha}
\gamma^+ u(\check{k},h)
\ea
with the Feynman propagator defined as
\ba
\label{eq:propagator}
S_F(x,y)_{\beta\alpha} &=& \langle 0 | \hat T [\Psi_\beta(x) \bar \Psi_\alpha(y)] | 0 \rangle
\ea
where $\hat T$ being the time ordering operator.
Finally,  the quark Feynman propagator in background field, $S_F(x,y)_{\alpha\beta}$, is derived in Sec.~\ref{sec:quark_prop} and its final form is written in Eq. \eqref{unpol_hdep_decomp} with the explicit expressions of the unpolarized piece given in Eq. \eqref{q_prop_ba_h_indep} and the helicity dependent piece given in Eq. \eqref{q_prop_ba_h_dep} at NEik accuracy.


The form of the $S$-matrix given in Eq.~\eqref{eq:S-matrix} enables one to compute scattering elements through a  time-ordered correlation function and constitutes therefore a LSZ-type reduction formula. To compute physical observables, such as the production  cross section, it is convenient to define the scattering amplitude
\ba
\mathcal{M}_{q(\check{q},h',\beta)\leftarrow q(\check{k},h,\alpha)} \equiv
\mathcal{M}^{hh'}_{\alpha\beta}(\uk,\q)
\ea
from the $S$-matrix, by extracting the trivial delta function as
\ba
\label{eq:S-matrix-M}
S_{ q(\check{q},h',\beta)\leftarrow q(\check{k},h,\alpha)} &=& (2k^+)2\pi \delta(q^+ - k^+)
i \mathcal{M}^{hh'}_{\alpha\beta}(\uk,\q) \, .
\ea

Inserting the expression of the Feynman propagator (Eq. \eqref{unpol_hdep_decomp} together with Eqs. \eqref{q_prop_ba_h_indep} and \eqref{q_prop_ba_h_dep})  into Eq.~\eqref{eq:S-matrix} and then comparing it to Eq.~\eqref{eq:S-matrix-M}, it is straightforward to extract the scattering amplitude at NEik accuracy as
\ba
\label{eq:amplitude-spinors}
i \mathcal{M}^{hh'}_{\alpha\beta}(\uk,\q)&=&
\frac{1}{2k^+}
\int d^2 \z\, e^{-i\z\cdot(\q\!-\!\k)} \, \bar u(\check{q},h') \gamma^+
\Bigg\{\mathcal{U}_F\Big(\frac{L^+}{2},-\frac{L^+}{2};\z\Big)
\\
&&
+\frac{i[\gamma^i,\gamma^j]}{8k^+}\int_{-\frac{L^+}{2}}^{\frac{L^+}{2}}dz^+\,
\left[\mathcal{U}_F\Big(\frac{L^+}{2},z^+;\z\Big)\,
\Big(-igt \!\cdot \!\mathcal{F}_{ij}(\uz) \Big)
\mathcal{U}_F\Big(z^+,-\frac{L^+}{2};\z\Big)\right]
\nn\\
&&
-\frac{i}{2k^+}\int_{-\frac{L^+}{2}}^{\frac{L^+}{2}}dz^+\,  \left[\mathcal{U}_F\Big(\frac{L^+}{2},z^+;\z\Big)\,
\overleftarrow{\mathcal{D}_{\z^j}}\, \overrightarrow{\mathcal{D}_{\z^j}}\,
\mathcal{U}_F\Big(z^+,-\frac{L^+}{2};\z\Big)\right]
\nn\\
&&
-\frac{(\q^j\!+\!\k^j)}{4k^+}\int_{-\frac{L^+}{2}}^{\frac{L^+}{2}}dz^+\, \left[\mathcal{U}_F\Big(\frac{L^+}{2},z^+;\z\Big)\,
\overleftrightarrow{\mathcal{D}_{\z^j}}\,
\mathcal{U}_F\Big(z^+,-\frac{L^+}{2};\z\Big)\right]
\Bigg\}_{\alpha\beta} u(\check{k},h) \nn
\ea
where we have used
\ba
\label{Spinor_algebra_1}
 \gamma^+(\slashed{\check{k}} +m)\gamma^+ &=& 2k^+\gamma^+\; , \\
\label{Spinor_algebra_2}
\gamma^+(\slashed{\check{q}} +m)\gamma^+ &=& 2q^+\gamma^+=2k^+ \gamma^+\; .
 \ea
Note that the last equality in Eq.~\eqref{Spinor_algebra_2} holds due to the presence of the delta function in Eq.~\eqref{eq:S-matrix-M}.  We would like to emphasize that the second term in Eq.~\eqref{eq:amplitude-spinors} which is proportional to $[\gamma^i,\gamma^j]$ introduces the helicity dependence on the scattering amplitude at NEik order and it governs the polarization effects during the scattering process. This helicity dependent term can be further simplified by using the helicity operator $S^3$, which acts on spinors as
\ba
S^3 u(\check{k},h) &=& h u(\check{k},h),\\
\label{S3-on-v}
S^3 v(\check{k},h) &=& -h v(\check{k},h).
\ea
Since $[\gamma^i,\gamma^j]=-4i\epsilon^{ij} S^3$, the spinor structure of the helicity dependent term in the scattering amplitude given in Eq.~\eqref{eq:amplitude-spinors} can be simplified to
\ba
\label{hel}
\bar u(\check{q},h')\gamma^+ \frac{[\gamma^i, \gamma^j]}{4} u(\check{k},h) =
-i\epsilon^{ij} h \; \bar u(\check{q},h') \gamma^+ u(\check{k},h)
\ea
where $\epsilon^{ij}$ is the antisymmetric tensor with $\epsilon^{12}=+1$. By using this simplification for the helicity dependent term and the relation in Eq.~\eqref{eq:condition-spinors}, the scattering amplitude can be written at NEik accuracy as
\ba
\label{eq:amplitude}
i \mathcal{M}^{hh'}_{\alpha\beta}(\uk,\q)&=&
\delta_{hh'}
\int d^2 \z\, e^{-i\z\cdot(\q\!-\!\k)} \,
\Bigg\{\mathcal{U}_F\Big(\frac{L^+}{2},-\frac{L^+}{2};\z\Big)
\nn \\
&&
+\frac{\epsilon^{ij} h}{2k^+}\int_{-\frac{L^+}{2}}^{\frac{L^+}{2}}dz^+\,
\left[\mathcal{U}_F\Big(\frac{L^+}{2},z^+;\z\Big)\,
\Big(-igt \!\cdot \!\mathcal{F}_{ij}(\uz) \Big)
\mathcal{U}_F\Big(z^+,-\frac{L^+}{2};\z\Big)\right]
\nn\\
&&
-\frac{i}{2k^+}\int_{-\frac{L^+}{2}}^{\frac{L^+}{2}}dz^+\,  \left[\mathcal{U}_F\Big(\frac{L^+}{2},z^+;\z\Big)\,
\overleftarrow{\mathcal{D}_{\z^j}}\, \overrightarrow{\mathcal{D}_{\z^j}}\,
\mathcal{U}_F\Big(z^+,-\frac{L^+}{2};\z\Big)\right]
\nn\\
&&
-\frac{(\q^j\!+\!\k^j)}{4k^+}\! \int_{-\frac{L^+}{2}}^{\frac{L^+}{2}}dz^+\, \left[\mathcal{U}_F\Big(\frac{L^+}{2},z^+;\z\Big)\,
\overleftrightarrow{\mathcal{D}_{\z^j}}\,
\mathcal{U}_F\Big(z^+,-\frac{L^+}{2};\z\Big)\right]
\Bigg\}_{\alpha\beta}  .
\ea
Eq.~\eqref{eq:amplitude} is the final form of the forward quark scattering amplitude at NEik accuracy.

\subsection{Unpolarized partonic cross section}
According to the hybrid factorization ansatz \cite{Dumitru:2005gt}, in order to compute the hadronic cross section one first computes the partonic cross section which in our case corresponds to projectile quark (or antiquark) scattering on the dense target. Then, the partonic cross section is convoluted with the quark distribution function in the proton and with the fragmentation function. However, here we aim for qualitative understanding of the effects of quark propagator computed at NEik accuracy. Therefore, we restrict ourselves to the computation of the partonic cross section and neglect the study of the convolution of it with the parton distribution functions and fragmentation functions.

In the hybrid factorization the incoming parton is treated in the collinear limit. This corresponds to describing the incoming parton with vanishing transverse momenta $\k$ which is necessary when convoluting the partonic cross section with the parton distribution functions.  Therefore, when computing the observables in the hybrid factorization at hadronic level one sets the incoming momenta of the parton $\k \to 0$.  However, since we are only interested in the qualitative analysis of the cross section at partonic level, we keep our expression of the quark scattering amplitude more general and avoid setting the transverse momenta of the incoming parton to zero.

At partonic level, the differential cross section of the quark traversing the target and undergoing multiple scatterings with the strong background field is given as
\ba
\label{eq:cross-section-quark}
d^3\sigma_{q(\check{k},h,\alpha)\rightarrow  q(\check{q},h',\beta)} = (2q^+) 2\pi \delta(q^+-k^+)
\mathcal{M}^{hh'}_{\alpha\beta}(\uk,\q)^\dagger \mathcal{M}^{hh'}_{\alpha\beta}(\uk,\q)
\frac{\theta(q^+)}{2q^+} \frac{d^2\q}{(2\pi)^2} \frac{dq^+}{2\pi}
\ea
with $\mathcal{M}^{hh'}_{\alpha\beta}(\uk,\q)$ being the forward quark scattering amplitude. After performing the trivial integration over the outgoing quark longitudinal momentum $q^+$, the partonic cross section can be cast into
\ba
\label{eq:dif-cross-section}
\frac{d^2\sigma^{qA\to q+X}}{d^2\q} 
&=& \frac{1}{(2\pi)^2}\, \frac{1}{2N_c} \sum_{h,h'} \sum_{\alpha,\beta}
\mathcal{M}^{hh'}_{\alpha\beta}(\uk,\q)^\dagger \mathcal{M}^{hh'}_{\alpha\beta}(\uk,\q) \bigg|_{q^+=k^+} \, .
\ea
Here, the factor $2$ in the denominator originates from averaging over the initial state quark helicity and the factor $N_c$ from averaging over the color, while summing over the final state helicity and color.

By using the explicit expression for the forward quark scattering amplitude given in Eq.~\eqref{eq:amplitude}, the partonic level production cross section at NEik accuracy can be written as
\ba
\label{eq:cross-section-unpol-explicit}
\frac{d^2\sigma^{qA\to q+X}}{d^2\q}
&=& \frac{1}{N_c} \frac{1}{(2\pi)^2}
\int d^2 \z' \int d^2 \z\,  e^{-i(\q-\k)\cdot(\z\!-\!\z')} \,
{\rm Tr}
\Bigg\{\mathcal{U}_F^\dagger (\z') \mathcal{U}_F(\z) \\
&&
\hspace{-2.2cm}
+\frac{1}{2k^+}\int_{-\frac{L^+}{2}}^{\frac{L^+}{2}}dz^+\,
\mathcal{U}_F^\dagger (\z' )\,
\mathcal{U}_F \Big(\frac{L^+}{2},z^+;\z\Big)\,
\left[ -i\overleftarrow{\mathcal{D}_{\z^j}}\, \overrightarrow{\mathcal{D}_{\z^j}}\,
-\frac{(\q^j+\k^j)}{2} \overleftrightarrow{\mathcal{D}_{\z^j}} \right]
\mathcal{U}_F\Big(z^+,-\frac{L^+}{2};\z\Big)
\nn\\
&&
\hspace{-2.2cm}
+\frac{1}{2k^+}\int_{-\frac{L^+}{2}}^{\frac{L^+}{2}}dz^+\,
\mathcal{U}_F^\dagger \Big(z^+,-\frac{L^+}{2};\z'\Big)\,
\left[ i\overleftarrow{\mathcal{D}_{\z'^j}}\, \overrightarrow{\mathcal{D}_{\z'^j}}\,
+\frac{(\q^j+\k^j)}{2} \overleftrightarrow{\mathcal{D}_{\z'^j}} \right]
\mathcal{U}_F^\dagger \Big(\frac{L^+}{2},z^+;\z'\Big)
\mathcal{U}_F (\z)
\Bigg\} \nn
\ea
in the case of unpolarized target and projectile. In order to arrive at Eq.~\eqref{eq:cross-section-unpol-explicit}, we performed the sums over the initial and final state helicities, introduced a simplified notation
$\mathcal{U}_F (\z) \equiv \mathcal{U}_F\Big(\frac{L^+}{2},-\frac{L^+}{2};\z\Big)$ and used the following identities for the covariant derivatives\footnote{Note that helicity dependent term in the forward quark scattering amplitude given in Eq.~\eqref{eq:amplitude} vanishes at the level of the cross section after summing over incoming and outgoing quark helicities.}:
\ba
\left[\mathbb{A}(\z)\, \overleftarrow{\mathcal{D}_{\z^j}}\, \mathbb{B}(\z)\right]^\dagger
&=& \mathbb{B}(\z)^\dagger\, \overrightarrow{\mathcal{D}_{\z^j}}\, \mathbb{A}(\z)^\dagger
\\
\left[\mathbb{A}(\z)\, \overrightarrow{\mathcal{D}_{\z^j}}\, \mathbb{B}(\z)\right]^\dagger
&=& \mathbb{B}(\z)^\dagger\,  \overleftarrow{\mathcal{D}_{\z^j}}\, \mathbb{A}(\z)^\dagger
\\
\left[\mathbb{A}(\z)\, \overleftrightarrow{\mathcal{D}_{\z^j}}\, \mathbb{B}(\z)\right]^\dagger
&=& -  \mathbb{B}(\z)^\dagger\,  \overleftrightarrow{\mathcal{D}_{\z^j}}\, \mathbb{A}(\z)^\dagger
\ea
where $\mathbb{A}(\z)$ and $\mathbb{B}(\z)$ are generic $SU(N_c)$ matrix-valued functions of $\z$.

Let us now analyze symmetry properties of the unpolarized cross section. Within the CGC framework, one needs to perform the target averaging of the traces of the Wilson lines that appear in Eq. \eqref{eq:cross-section-unpol-explicit}. After introducing the target averaging,  the unpolarized cross section reads
\ba
\bigg\langle \frac{d^2\sigma^{qA\to q+X}}{d^2\q}\bigg\rangle_A
 &=&\frac{1}{(2\pi)^2}\int d^2\r \; e^{-i(\q-\k)\cdot\r}\bigg\{ d_F(\r)-\frac{1}{2k^+}\frac{(\q^j+\k^j)}{2}\Big[ {\cal O}_{(1)}^j(\r)+{\cal O}_{(1)}^{\dagger j}(-\r)\Big]
 \nonumber\\
&&
\hspace{5.2cm}
-\frac{i}{2k^+}\Big[ {\cal O}_{(2)}(\r)-{\cal O}^\dagger_{(2)}(-\r)\Big]\bigg\}
\ea
where we have introduced the new integration variables $(\z-\z')\equiv \r$ and  $(\z+\z')\equiv 2\b$. Moreover, we have also defined the following  dipole and decorated dipole operators:
\ba
\label{dipole}
d_F(\r)&=&\frac{1}{N_c} \int d^2\b \; \bigg\langle {\rm tr}\bigg[ \mathcal{U}^\dagger_F\Big(\frac{L^+}{2},-\frac{L^+}{2};\b-\frac{\r}{2}\Big)\, \mathcal{U}_F\Big(\frac{L^+}{2},-\frac{L^+}{2};\b+\frac{\r}{2}\Big) \bigg]\bigg\rangle_A  \, ,\\
\label{decorated-dipole-1}
{\cal O}_{(1)}^j(\r)&=&\frac{1}{N_c} \int d^2\b \; \int_{-\frac{L^+}{2}}^{\frac{L^+}{2}} dz^+ \, \bigg\langle {\rm tr}\bigg[ {\mathcal U}^\dagger_F\Big( \b-\frac{\r}{2}\Big)
\\
&&\hspace{3.5cm}
\times\;
 {\mathcal U}_F\Big(\frac{L^+}{2},z^+; \b+\frac{\r}{2}\Big) \overleftrightarrow{\mathcal{D}_{\b^j+\frac{\r^j}{2}}} {\mathcal U}_F\Big( z^+,-\frac{L^+}{2}; \b+\frac{\r}{2}\Big)\bigg]\bigg\rangle_A \, ,\nonumber\\
 \label{decorated-dipole-2}
{\cal O}_{(2)}(\r)&= &\frac{1}{N_c} \; \int d^2\b \;\int_{-L^+/2}^{L^+/2} dz^+ \, \bigg\langle {\rm tr}\bigg[ {\mathcal U}^\dagger_F\Big( \b-\frac{\r}{2}\Big)
\\
&&\hspace{2.1cm}
\times\; {\mathcal U}_F\Big(\frac{L^+}{2},z^+; \b+\frac{\r}{2}\Big)\overleftarrow{\mathcal{D}_{\b^j+\frac{\r^j}{2}}}\, \overrightarrow{\mathcal{D}_{\b^j+\frac{\r^j}{2}}} {\mathcal U}_F\Big( z^+,-\frac{L^+}{2}; \b+\frac{\r}{2}\Big)\bigg]\bigg\rangle_A \, .\nonumber
\ea
The target averaging procedure and the integration over the impact parameter $\b$ ensure the covariance of the operators under rotations in the transverse plane. In particular, the operator ${\cal O}_{(1)}^j(\r)$ behaves as a vector quantity under rotations in the transverse plane, and therefore satisfies
\ba
\label{O1_sym}
{\cal O}^{ j}_{(1)}(-\r) = - {\cal O}^{j}_{(1)}(\r)\\
{\cal O}^{\dagger j}_{(1)}(-\r) = - {\cal O}^{\dagger j}_{(1)}(\r)
\ea
On the contrary, the operator ${\cal O}_{(2)}(\r)$ behaves as a scalar quantity under rotations in the transverse plane, so that
\ba
\label{O2_sym}
{\cal O}_{(2)}(-\r) = {\cal O}_{(2)}(\r)\\
{\cal O}^{\dagger}_{(2)}(-\r) = {\cal O}^{\dagger }_{(2)}(\r)
\ea
Thus, using the above symmetry properties, target-averaged unpolarized cross section can be written as
\ba
\bigg\langle \frac{d^2\sigma^{qA\to q+X}}{d^2\q} \bigg\rangle_A&=&\frac{1}{(2\pi)^2}\int d^2\r \; e^{-i(\q-\k)\cdot\r}\bigg\{ d_F(\r)-\frac{1}{2k^+}\frac{(\q^j+\k^j)}{2}\Big[ {\cal O}_{(1)}^j(\r)-{\cal O}_{(1)}^{\dagger j}(\r)\Big]
 \nonumber\\
&&
\hspace{5.1cm}
-\frac{i}{2k^+}\Big[ {\cal O}_{(2)}(\r)-{\cal O}^\dagger_{(2)}(\r)\Big]\bigg\}
\ea

Let us finally comment on the structure of the NEik corrections to the unpolarized cross section. The strict eikonal term which is a fundamental dipole can be decomposed into Pomeron and Odderon operators. These operators are defined in terms of the fundamental dipole and its hermitian conjugate:
\ba
\label{Pomeron}
P(\x,\y)&=& \frac{1}{2}\bigg[ 2 - d_F(\x,\y)-d^\dagger_F(\x,\y)\bigg]\\
\label{Odderon}
O(\x,\y)&=&\frac{1}{2}\bigg[ d_F(\x,\y)-d^\dagger_F(\x,\y)\bigg]
\ea
with the fundamental dipole operator defined in the standard way.
In order to preserve the consistency, we introduce the same change of variables ($\r$ and $\b$)
in the definition of the Pomeron and Odderon operators (Eqs. \eqref{Pomeron} and \eqref{Odderon} respectively), and define the following new operators which are integrated expressions of the Pomeron and Odderon operators over $\b$.
\ba
\label{bar-Pomeron}
{\bar P}(\r)&\equiv&\int d^2\b \; P\Big( \b+\frac{\r}{2},\b-\frac{\r}{2}\Big)= \frac{1}{2}\Big[ 2-d_F(\r)-d^\dagger_F(\r)\Big] \, \\
\label{bar-Odderon}
{\bar O}(\r) &\equiv&\int d^2\b \; O\Big( \b+\frac{\r}{2},\b-\frac{\r}{2}\Big)= \frac{1}{2}\Big[ d_F(\r)-d^\dagger_F(\r)\Big] \, .
\ea
By using these definitions, the target-averaged unpolarized partonic cross section for a quark can be written as
\ba
\label{final-quark-unpolarized}
\bigg\langle \frac{d^2\sigma^{qA\to q+X}}{d^2\q} \bigg\rangle_A&=&\frac{1}{(2\pi)^2}\int d^2\r \; e^{-i(\q-\k)\cdot\r}\bigg\{ 1- \bar P(\r) \nonumber\\
&&
\hspace{-0.9cm}
+\bigg\lgroup \bar O(\r)
-\frac{1}{2k^+}\frac{(\q^j+\k^j)}{2}\Big[ {\cal O}_{(1)}^j(\r)-{\cal O}_{(1)}^{\dagger j}(\r)\Big]
-\frac{i}{2k^+}\Big[ {\cal O}_{(2)}(\r)-{\cal O}^\dagger_{(2)}(\r)\Big]\bigg\rgroup\bigg\}\nonumber\\
&=&
\frac{1}{(2\pi)^2}\int d^2\r \;
\bigg\{ \cos\left[(\q-\k)\!\cdot\! \r\right]\,\Big[1- \bar P(\r)\Big]
\\
&&
\hspace{-2.3cm}
+\sin\left[(\q-\k)\!\cdot\!\r\right]\,
\Big[\textrm{Im}\bar O(\r) - \frac{(\q^j+\k^j)}{2k^+}\, \textrm{Im}{\cal O}_{(1)}^j(\r)\Big]
+ \cos\left[(\q-\k)\!\cdot\!\r\right]\,
\frac{1}{k^+}\, \textrm{Im}{\cal O}_{(2)}(\r)
\bigg\} \, .\nonumber
\ea

The analogous expression for antiquark-nucleus scattering can be calculated in a similar way (see Appendix \ref{Sec:antiquark-observables} for the details of the calculation). The final result reads
\ba
\label{final-antiquark-unpolarized}
\bigg\langle \frac{d^2\sigma^{\bar qA\to \bar q+X}}{d^2\q} \bigg\rangle_A&=&\frac{1}{(2\pi)^2}\int d^2\r \; e^{-i(\q-\k)\cdot\r}\bigg\{ 1- \bar P(\r) \nonumber\\
&&
\hspace{-0.9cm}
-\bigg\lgroup
\bar O(\r)
-\frac{1}{2k^+}\frac{(\q^j+\k^j)}{2}\Big[ {\cal O}_{(1)}^j(\r)-{\cal O}_{(1)}^{\dagger j}(\r)\Big]
-\frac{i}{2k^+}\Big[ {\cal O}_{(2)}(\r)-{\cal O}^\dagger_{(2)}(\r)\Big]\bigg\rgroup\bigg\}\nonumber\\
&=&
\frac{1}{(2\pi)^2}\int d^2\r \;
\bigg\{ \cos\left[(\q-\k)\!\cdot\r\right]\,\Big[1- \bar P(\r)\Big]
\\
&&
\hspace{-2.3cm}
-\sin\left[(\q-\k)\!\cdot\r\right]\,
\Big[\textrm{Im}\bar O(\r) - \frac{(\q^j+\k^j)}{2k^+}\, \textrm{Im}{\cal O}_{(1)}^j(\r)\Big]
- \cos\left[(\q-\k)\!\cdot\r\right]\,
\frac{1}{k^+}\, \textrm{Im}{\cal O}_{(2)}(\r)
\bigg\}\, . \nonumber
\ea

As it is well known, Pomeron is even under both signature transformation (${\mathcal U}(\x)\to {\mathcal U}^\dagger(\x)$) and charge conjugation (quark $\to$ antiquark) while Odderon is odd under both transformations. By comparing Eqs. \eqref{final-quark-unpolarized} and \eqref{final-antiquark-unpolarized}, it is straightforward to realize that
NEik corrections in the unpolarized partonic cross section are also odd under signature transformation and charge conjugation. Therefore, one can conclude that the NEik corrections to the unpolarized partonic cross section couple to Odderon. We would like to mention that for the unpolarized cross section, Pomeron type contribution is seen at the strict eikonal order. However, we also expect to see such contributions at NNEik accuracy just like the analogue contributions for gluon propagator computed in \cite{Altinoluk:2015gia}. On the other hand, the Odderon type contributions are seen both at eikonal and NEik orders.

\subsection{Quark helicity asymmetry}

The other application considered here for our background quark propagator at NEik accuracy concerns the quark helicity asymmetry. In the analysis of this quantity, we again restrict ourselves to the partonic level, meaning quark scattering on a nucleus, and aim at gaining some qualitative insight on the effects of NEik corrections on asymmetric quantities such as quark helicity asymmetry.

We consider the difference between the cross sections for a quark of positive and negative helicity  scattering on the nucleus target
\ba
\frac{d^2\Delta\sigma^{ qA\to  q+X}}{d^2\q}\equiv \frac{d^2\sigma^{ q^{(+)}A\to  q+X}}{d^2\q} -\frac{d^2\sigma^{ q^{(-)}A\to  q+X}}{d^2\q}
\ea
that we call as the quark helicity asymmetry.\footnote{In principle an asymmetry is defined as the ratio of difference over the sum of positive and negative helicity contributions. Here, we focus on the numerator for simplicity and still call it as asymmetry with a small abuse of language.}
This asymmetry,  when written in terms of the forward quark scattering amplitude $\mathcal{M}^{hh'}_{\alpha\beta}(\uk,\q)$ (keeping the conventions from Fig.~\ref{fig:quark}), reads
\ba
\label{eq:dif-cross-section-pol}
\frac{d^2\Delta\sigma^{ qA\to  q+X}}{d^2\q}=\frac{1}{(2\pi)^2}\, \frac{1}{2N_c} \sum_{h,h'} \sum_{\alpha,\beta} (2h)\,
\mathcal{M}^{hh'}_{\alpha\beta}(\uk,\q)^\dagger \mathcal{M}^{hh'}_{\alpha\beta}(\uk,\q) \bigg|_{q^+=k^+}
\ea
since $h=\pm 1/2$.
After inserting the definition of the forward quark scattering amplitude, given in Eq.~\eqref{eq:amplitude}, into the definition of the quark helicity asymmetry \eqref{eq:dif-cross-section-pol}, we get
\ba
\label{eq:cross-section-pol-explicit}
\frac{d^2\Delta\sigma^{ qA\to  q+X}}{d^2\q}
&=& \frac{1}{N_c} \frac{1}{(2\pi)^2}
\int d^2 \z' \int d^2 \z\,  e^{-i(\q-\k)\cdot(\z\!-\!\z')} \, \frac{1}{4k^+}\\
&&
\hspace{-1.4cm}
\times \, {\rm tr} \,
\bigg\{\int_{-\frac{L^+}{2}}^{\frac{L^+}{2}}dz^+\,
\mathcal{U}_F^\dagger (\z' )\,
\mathcal{U}_F \Big(\frac{L^+}{2},z^+;\z\Big)\,
\big[\epsilon^{ij}\left( -i g t\! \cdot \! \mathcal{F}_{ij}(\uz) \right)\big]
\mathcal{U}_F\Big(z^+,-\frac{L^+}{2};\z\Big)
\nn\\
&&
\hspace{-1cm}
+\int_{-\frac{L^+}{2}}^{\frac{L^+}{2}}dz^+\,
\mathcal{U}_F^\dagger \Big(z^+,-\frac{L^+}{2};\z'\Big)\,
\big[\epsilon^{ij}\left( i g t\! \cdot \! \mathcal{F}_{ij}(z^+,\z') \right)\big]
\mathcal{U}_F^\dagger \Big(\frac{L^+}{2},z^+;\z'\Big)
\mathcal{U}_F (\z)
\Bigg\} \, .\nn
\ea
Before we continue with the symmetry properties of the quark helicity asymmetry, we would like to mention that, contrary to the unpolarized partonic cross, only the helicity dependent piece of the quark scattering amplitude contributes to the computation of the quark helicity asymmetry. Moreover, the unpolarized partonic cross section starts at eikonal order and gets correction at NEik order, while the quark helicity asymmetry starts at NEik order.

Let us now turn to the symmetry properties of the quark helicity asymmetry. We follow the same strategy that was introduced in the analysis of the symmetry properties of the unpolarized partonic cross section. We change the integration variables  to $\b$ and $\r$, and introduce the target averaging so that the quark helicity asymmetry can be written as
\ba
\Big\langle \frac{d^2\Delta\sigma^{ qA\to  q+X}}{d^2\q} \Big\rangle_A =
 \frac{1}{(2\pi)^2}\int d^2\r \, e^{-i(\q-\k)\cdot\r} \frac{(-i)}{4k^+} \Big[ O_{(3)}(\r)-O_{(3)}^\dagger(-\r)\Big]
\ea
where we have defined the new decorated dipole operator $O_{(3)}(\r)$ as
\ba
\label{O3}
O_{(3)}(\r)&=&\frac{1}{N_c}\int d^2\b \int_{-L^+/2}^{L^+/2}dz^+\bigg\langle {\rm Tr}\bigg[ {\cal U}^\dagger_F\Big(\b-\frac{\r}{2}\Big)\\
&&
\times \;
{\cal U}_F\Big(\frac{L^+}{2},z^+; \b+\frac{\r}{2}\Big)\Big\{ \epsilon^{ij}\Big[ gt\cdot {\cal F}_{ij}\Big(z^+, \b+\frac{\r}{2}\Big)\Big] \Big\} \,
{\cal U}_F\Big(z^+, -\frac{L^+}{2}; \b+\frac{\r}{2}\Big)\bigg]\bigg\rangle_A \, .\nonumber
\ea
Since in the newly defined operator $O_{(3)}(\r)$, the antisymmetric tensor $\epsilon^{ij}$ is already contracted with the $ij$ component of the field strength tensor, this operator $O_{(3)}(\r)$ behaves as a scalar quantity under rotations within the transverse plane, and therefore
\ba
\label{sym-O3}
O_{(3)}(-\r)=O_{(3)}(\r)\\
O^\dagger_{(3)}(-\r)=O^\dagger_{(3)}(\r)
\ea
and the target-averaged quark helicity asymmetry reads
\ba
\label{final-quark-helicity}
\Big\langle \frac{d^2\Delta\sigma^{ qA\to  q+X}}{d^2\q} \Big\rangle_A
&=&
 \frac{1}{(2\pi)^2}\int d^2\r \, e^{-i(\q-\k)\cdot\r} \frac{(-i)}{4k^+} \Big[ O_{(3)}(\r)-O_{(3)}^\dagger(\r)\Big]
\nonumber\\
&=&
 \frac{1}{(2\pi)^2}\,  \frac{1}{2k^+} \int d^2\r \; \cos\left[(\q-\k)\!\cdot\!\r\right]\;
 \textrm{Im} O_{(3)}(\r) \, .
\ea
The antiquark helicity asymmetry (see Appendix \ref{Sec:antiquark-observables} for the details of the calculation) leads to the same result at NEik  accuracy
\ba
\label{final-antiquark-helicity}
\Big\langle \frac{d^2\Delta\sigma^{ \bar{q}A\to  \bar{q}+X}}{d^2\q} \Big\rangle_A
&=&
 \frac{1}{(2\pi)^2}\int d^2\r \, e^{-i(\q-\k) \cdot \r} \frac{(-i)}{4k^+} \Big[ O_{(3)}(\r)-O_{(3)}^\dagger(\r)\Big]
 \nonumber\\
&=&
 \frac{1}{(2\pi)^2}\,  \frac{1}{2k^+} \int d^2\r \; \cos\left[(\q-\k)\!\cdot\!\r\right]\;
 \textrm{Im} O_{(3)}(\r) \, .
\ea

The form of the target-averaged helicity asymmetries given in Eqs. \eqref{final-quark-helicity} and \eqref{final-antiquark-helicity} suggests that it is odd  under signature transformation but even under charge conjugation which is compatible  neither with Pomeron nor Odderon type of behavior. These quantum numbers are instead observed in higher order Reggeons \cite{Altinoluk:2013rua, Altinoluk:2014twa}.

Let us look more closely into $O_{(3)}(\r)$ and the target average it contains, see Eq.~\eqref{O3}. One could pull the $\epsilon^{ij}$ out of the target average. The result of the target averaging should then be an antisymmetric transverse tensor in $ij$. This is not possible if it depends on a single vector $\r$. For that reason, $O_{(3)}(\r)$ vanishes in the case of an unpolarized target. Actually, $O_{(3)}(\r)$ involves $\epsilon^{ij}{\cal F}^a_{ij} = 2 {\cal F}^a_{12}$, which is the chromomagnetic field along the longitudinal axis, expected to vanish for an unpolarized target. By contrast, a target with a fixed non-zero helicity could support such a longitudinal chromomagnetic field, which is expected to flip sign if the target helicity is flipped. In conclusion, the expressions \eqref{final-quark-helicity} and \eqref{final-antiquark-helicity} are expected to vanish in the case of single longitudinal spin asymmetry, taking the difference of the two quark helicity contributions for the scattering on an unpolarized target. But the expressions \eqref{final-quark-helicity} and \eqref{final-antiquark-helicity} are expected to provide a non-vanishing result in the case of double longitudinal spin asymmetry, meaning taking the difference between the cases of same sign and opposite signs for the helicity of the quark projectile and the helicity of the target.

\section{Summary and outlook}
\label{sec:Discussions}
In conclusion, in this paper we derive the full NEik order corrections for both quark and antiquark propagators through a gluon background field. These corrections include the finite-width target effects which originate from relaxing the eikonal approximation that treats the dense target as an infinitely thin shockwave. Our derivation of the effects of the quark's transverse motion in the finite-width target 
treats the incoming and outgoing quark states in a symmetric manner, which differs from the one performed in Ref. \cite{Altinoluk:2014oxa} for a gluon propagator. Moreover, in this paper we have also included the NEik effects that stem from including the interactions of the quark propagator with the transverse component of the background field of the target which is neglected in the strict eikonal treatment.  The quark and antiquark propagators  that are derived at NEik accuracy are written as helicity dependent and helicity independent parts (given in Eqs. \eqref{q_prop_ba_h_dep} and \eqref{q_prop_ba_h_indep} for quarks and \eqref{qbar_prop_ba_h_dep}  and \eqref{qbar_prop_ba_h_indep} for antiquarks) which contribute to different observables. The structure of the NEik corrections to the quark propagator include covariant derivatives acting on the Wilson lines and therefore the results are gauge invariant (covariant). The effects of the interaction of the quark propagator with the transverse component of the background field of the target are crucial in order to arrive to a gauge invariant result. This was conjectured in \cite{Hatta:2016aoc} in a different context, but in this paper a step-by-step derivation is provided on how to reach a gauge invariant expression at NEik order.

We then apply the results to forward quark (or antiquark) nucleus collisions. The first application that we studied here is the unpolarized production cross section and we computed the NEik corrections to leading order quark (and antiquark) production cross section at forward rapidity. Since our main motivation is to gain qualitative insight on the effects of NEik corrections, we have restricted ourselves to study these unpolarized cross section at partonic level for simplicity instead of considering realistic proton-nucleus collisions. In this application, we have seen that only the helicity independent part of the quark (antiquark) part of the scattering amplitude contributes to the unpolarized cross section. Moreover, by comparing the resulting expressions for the quark and antiquark cross sections, we found that the NEik corrections are odd under both charge conjugation and signature transformations, thus exhibit  Odderon type behavior. The next application we considered is the quark (antiquark) helicity asymmetry. Contrary to the unpolarized cross section, only the helicity dependent part of the quark (antiquark) scattering amplitude contributes to the quark (antiquark) helicity asymmetry and it starts at NEik order. The symmetry properties of these helicity asymmetries have shown that they are even under charge conjugation and odd under signature transformation. Therefore, they behave  neither as Pomeron nor Odderon, but instead some higher Reggeon operator.

There are other possible applications of the results that are presented in this paper. The immediate application that we are planing to address is the NEik corrections to inclusive and diffractive dijet production in DIS. The NEik corrections to these observables are expected to be important for the future EIC since the scattering energies in this new machine will not be extremely high. Moreover, one can probe the small-x limit of the Weizs\"acker-Williams gluon TMDs in the so called correlation limit of this observable \cite{Dominguez:2011wm}. In this context, we plan to study the correlation limit at NEik accuracy and investigate the possibility of interpreting the NEik corrections as the corrections to the Weizs\"acker-Williams gluon TMDs. Moreover, it would be also interesting to search for a possible connection between the higher order kinematic twist contributions introduced in \cite{Altinoluk:2019fui, Altinoluk:2019wyu} and NEik corrections.

On the other hand, the angular correlations between hadrons and prompt photon are suggested to provide insight into the gluon fields of the target in proton-nucleus collisions. At forward rapidities\footnote{At central rapidities, the photon-jet angular correlations are studied in \cite{Benic:2016uku, Benic:2017znu, Benic:2018hvb} beyond the tree level at eikonal accuracy in proton-nucleus collisions.}, this observable is studied through $q\to q\gamma$ process at partonic level and it is argued to provide the dominant contribution in the fragmentation region of the projectile \cite{Gelis:2002ki, JalilianMarian:2008iz, JalilianMarian:2012bd}. Thus, it is interesting to consider the angular correlations in photon-jet forward production and study the effects of sub-eikonal contributions on the near and away side peaks.

As discussed previously, the NEik gluon propagator that is derived in \cite{Altinoluk:2014oxa}, accounts only for the transverse motion effects and neglects the contributions from the interaction of the gluon propagator with the transverse component of the background field of the target. We plan to revisit this derivation to include the missing contribution and derive the complete NEik order gluon propagator. Together with the NEik corrections to the quark propagator derived in this paper, the complete gluon propagator will provide us the building blocks to study the angular correlations between dijets at central rapidities beyond eikonal accuracy in pA collisions.

At a more fundamental level, a dedicated study of the apparent differences and possible disagreements between the results derived in this manuscript and the ones derived in \cite{Chirilli:2018kkw} for the quark propagator in the gluon background at NEik accuracy is in progress \cite{Altinoluk_Beuf}.

\acknowledgments

We thank Giovanni Antonio Chirilli and Yuri Kovchegov for useful remarks. TA is supported by Grant No. 2018/31/D/ST2/00666 (SONA\-TA 14 - National Science Centre, Poland). AC is partially supported by the National Science
Centre, Poland under grant 2018/29/B/ST2/00646. This work has been performed in the framework of COST Action CA 15213 ``Theory of hot matter and relativistic heavy-ion collisions" (THOR), MSCA RISE 823947 ``Heavy ion collisions: collectivity and precision in saturation physics''  (HI\-EIC) and has received funding from the European Un\-ion's Horizon 2020 research and innovation programme under grant agreement No. 824093.



\appendix
\section{Derivation of the NEik corrections to the quark propagator in a pure $\mathcal{A}^-$ background\label{appendix}}

Let us start with the general expression \eqref{q_prop_pure_A_minus_Fourier_2} for the correction to the quark propagator in a pure $\mathcal{A}^-$ background field. Fourier transforming it to position space (following the convention \eqref{q_prop_Fourier}), and performing the integrations over $\p_n$ (for $n$ from $1$ to $N\!-\!1$), and over $q^-$ and $k^-$, one finds
\begin{align}
\delta S_F&(x,y) \bigg|_{\textrm{pure }\mathcal{A}^{-}}
=
\int \frac{d^3\uq}{(2\pi)^3} \int \frac{d^3\uk}{(2\pi)^3}\, 2\pi \delta(q^+\!-\!k^+)\,
e^{-i x\cdot \check{q}}\, e^{i y\cdot \check{k}}\, \frac{(\slashed{\check{q}}+m)\gamma^+(\slashed{\check{k}}+m)}{(2k^+)^2}
\,
\nn\\
& \times\:
\sum_{N=1}^{+\infty}\int\left[\prod_{n=1}^{N}  d^3\underline{z_n}\right]\;
e^{i \underline{z_N}\cdot \check{q}}\, e^{-i \underline{z_1}\cdot \check{k}}\:
\left\{ \mathcal{P}_n \prod_{n=1}^{N} \left[-igt\!\cdot\!\mathcal{A}^-(\underline{z_n})\right]
\right\}\:
\nn\\
& \times\:
\left\{ \prod_{n=1}^{N-1} \left[   \frac{(-i)k^+}{2\pi(z_{n+1}^+\!-\!z_{n}^+)}\:
                e^{-i\frac{(z_{n+1}^+\!-\!z_{n}^+)m^2}{2k^+}}\:
                e^{i \frac{k^+(\z_{n+1}-\z_{n})^2}{2(z_{n+1}^+\!-\!z_{n}^+)}}
\right]\right\}
\nn\\
& \times\:
\left\{ \theta(k^+) \left(\prod_{n=0}^{N}\theta(z_{n+1}^+\!-\!z_{n}^+)\right)
+(-1)^{N+1}\,\theta(-k^+) \left(\prod_{n=0}^{N}\theta(z_{n}^+\!-\!z_{n+1}^+)\right) \right\}\;
\, ,
\label{q_prop_pure_A_minus_1}
\end{align}
using the notations $z_{0}^+\equiv y^+$ and $z_{N+1}^+\equiv x^+$, and keeping fundamental color indices implicit from now on.
It is then convenient to simplify the phase factors by performing the change of variables $\z_{n}\mapsto \u_{n}$, defined by
\begin{align}
\z_{n} & \equiv  \u_{n} + \frac{1}{2} \left(\frac{\q}{q^+}+\frac{\k}{k^+}\right)z_{n}^+ = \u_{n} + \frac{z_{n}^+(\q\!+\!\k)}{2k^+}
\, .
\label{zn_to_un_change_of_var}
\end{align}
This can be interpreted as follows. In the $2+1$ dimensional Galilean subgroup of the Poincar\'e group in light-cone coordinates, $\p/p^+$ plays the role of a transverse Galilean velocity, and $z^+$ the role of time. Hence, the change of variables \eqref{zn_to_un_change_of_var} amounts to express the successive transverse positions of the quark in a transversely moving frame, with a Galilean velocity which is the average between the initial velocity of the quark $\k/k^+$ and the final one $\q/q^+$. Eq.~\eqref{q_prop_pure_A_minus_1} then becomes
\begin{align}
&\delta S_F(x,y) \bigg|_{\textrm{pure }\mathcal{A}^{-}}
=
\int \frac{d^3\uq}{(2\pi)^3} \int \frac{d^3\uk}{(2\pi)^3}\, 2\pi \delta(q^+\!-\!k^+)\,
e^{-i x\cdot \check{q}}\, e^{i y\cdot \check{k}}\, \frac{(\slashed{\check{q}}+m)\gamma^+(\slashed{\check{k}}+m)}{(2k^+)^2}
\,
\nn\\
& \times\:
\sum_{N=1}^{+\infty}\int\left[\prod_{n=1}^{N}  dz_n^+\right]
\left\{ \theta(k^+) \left(\prod_{n=0}^{N}\theta(z_{n+1}^+\!-\!z_{n}^+)\right)
+(-1)^{N+1}\,\theta(-k^+) \left(\prod_{n=0}^{N}\theta(z_{n}^+\!-\!z_{n+1}^+)\right) \right\}\;
\nn\\
& \times\:
e^{i(z_{N}^+-z_{1}^+) \frac{(\q-\k)^2}{8k^+}}
\int\left[\prod_{n=1}^{N}  d^2\u_n\right]\;
 e^{-\frac{i}{2}(\u_N+\u_1)\cdot (\q-\k)}
\left\{ \prod_{n=1}^{N-1} \left[   \frac{(-i)k^+}{2\pi(z_{n+1}^+\!-\!z_{n}^+)}\:
                e^{i \frac{k^+(\u_{n+1}-\u_{n})^2}{2(z_{n+1}^+\!-\!z_{n}^+)}}
\right]\right\}
\nn\\
& \times\:
\left\{ \mathcal{P}_n \prod_{n=1}^{N} \left[-igt\!\cdot\!\mathcal{A}^-\left(z_n^+,\u_{n} + \frac{z_{n}^+(\q\!+\!\k)}{2k^+}\right)\right]
\right\}\:
\, .
\label{q_prop_pure_A_minus_2}
\end{align}
The next step is to perform the change of variables from $\u_{n}$ (for $n$ from $1$ to $N$) to $\w_{n}$ (for $n$ from $1$ to $N-1$) and $\z$, defined as\footnote{Note that this change of variables has a unit Jacobian, for any $N\geqslant 1$.}
\begin{align}
\w_{n} & \equiv  \u_{n+1}-\u_{n}
\nn\\
\z & \equiv \frac{(\u_{N}+\u_{1})}{2}
\, .
\label{un_to_wn_and_z_change_of_var}
\end{align}
Then, one finds from Eq.~\eqref{q_prop_pure_A_minus_2}
\begin{align}
&\delta S_F(x,y) \bigg|_{\textrm{pure }\mathcal{A}^{-}}
=
\int \frac{d^3\uq}{(2\pi)^3} \int \frac{d^3\uk}{(2\pi)^3}\, 2\pi \delta(q^+\!-\!k^+)\,
e^{-i x\cdot \check{q}}\, e^{i y\cdot \check{k}}\, \frac{(\slashed{\check{q}}+m)\gamma^+(\slashed{\check{k}}+m)}{(2k^+)^2}
\,
\nn\\
& \times\:
\sum_{N=1}^{+\infty}\int\left[\prod_{n=1}^{N}  dz_n^+\right]
\left\{ \theta(k^+) \left(\prod_{n=0}^{N}\theta(z_{n+1}^+\!-\!z_{n}^+)\right)
+(-1)^{N+1}\,\theta(-k^+) \left(\prod_{n=0}^{N}\theta(z_{n}^+\!-\!z_{n+1}^+)\right) \right\}\;
\nn\\
& \times\:
e^{i(z_{N}^+-z_{1}^+) \frac{(\q-\k)^2}{8k^+}}
\int d^2\z\, e^{-i \z \cdot (\q-\k)}
\;\; {\cal E}
\, ,
\label{q_prop_pure_A_minus_3}
\end{align}
where
\begin{align}
{\cal E}  \equiv &
\int\left\{ \prod_{n=1}^{N-1}  \left[d^2\w_n\;
  \frac{(-i)k^+}{2\pi(z_{n+1}^+\!-\!z_{n}^+)}\:
                e^{i \frac{k^+\, \w_{n}^2}{2(z_{n+1}^+\!-\!z_{n}^+)}}
\right]\right\}
\nn\\
& \times\:
\left\{ \mathcal{P}_n \prod_{n=1}^{N} \left[-igt\!\cdot\!\mathcal{A}^-\left(z_n^+,\hat{\z}_{n}
+\frac{1}{2}\sum_{n'=1}^{n-1}\w_{n'}  - \frac{1}{2}\sum_{n'=n}^{N-1}\w_{n'}
\right)\right]
\right\}\:
\, ,
\label{q_prop_pure_A_minus_3bis}
\end{align}
and using the notation $\hat{\z}_{n}\equiv \z +z_n^+(\q+\k)/(2k^+)$ for the analog of the classical trajectory used in Refs.~\cite{Altinoluk:2014oxa,Altinoluk:2015gia}. The Taylor expansion of each background field insertion around its classical position $\hat{\z}_{n}$ allows us to go systematically beyond the Eikonal approximation, and to perform the integrations over $\w_{n}$ analytically thanks to
\begin{align}
\int d^2\w_n\;  \frac{(-i)k^+}{2\pi(z_{n+1}^+\!-\!z_{n}^+)}\;
                e^{i \frac{k^+\, \w_{n}^2}{2(z_{n+1}^+\!-\!z_{n}^+)}}
& = 1
\label{wn_int_0}\\
\int d^2\w_n\;  \frac{(-i)k^+}{2\pi(z_{n+1}^+\!-\!z_{n}^+)}\;
                e^{i \frac{k^+\, \w_{n}^2}{2(z_{n+1}^+\!-\!z_{n}^+)}}\: \w_n^i\,
& = 0
\label{wn_int_1}\\
\int d^2\w_n\;  \frac{(-i)k^+}{2\pi(z_{n+1}^+\!-\!z_{n}^+)}\;
                e^{i \frac{k^+\, \w_{n}^2}{2(z_{n+1}^+\!-\!z_{n}^+)}}\: \w_n^i\, \w_n^j\,
& =
\frac{i(z_{n+1}^+\!-\!z_{n}^+)}{k^+}\; \delta^{ij}
\label{wn_int_2}
\, .
\end{align}
In order to reach Next-to Eikonal accuracy, we have thus to account for all contributions which are quadratic in $\w_n$ in the Taylor expansion, for each $n$ separately. Note that the two powers of $\w_n$ do not necessarily come from the Taylor expansion of the same $\mathcal{A}^-$ insertion. After some algebra, we find in this way
\begin{align}
{\cal E}  = & \left\{ \mathcal{P}_n \prod_{n=1}^{N} \left[-igt\!\cdot\!\mathcal{A}^-\left(z_n^+,\hat{\z}_{n}\right)\right]\right\}
\nn\\
&
+\frac{i(z_{N}^+\!-\!z_{1}^+)}{8k^+}\, \sum_{n_1=1}^{N}\left\{ \mathcal{P}_n
\left[-igt\!\cdot\!\d_j \d_j\mathcal{A}^-\left(z_{n_1}^+,\hat{\z}_{n_1}\right)\right]
\prod_{  \begin{array}{c}
           n=1 \\
           n\neq n_1 \\
         \end{array}
}^{N} \left[-igt\!\cdot\!\mathcal{A}^-\left(z_n^+,\hat{\z}_{n}\right)\right]\right\}
\nn\\
&
+\frac{i}{4k^+}\, \sum_{n_1=1}^{N-1}\; \sum_{n_2=n_1+1}^{N} \Big[(z_{N}^+\!-\!z_{n_2}^+)-(z_{n_2}^+\!-\!z_{n_1}^+)+(z_{n_1}^+\!-\!z_{1}^+)\Big]
\nn\\
& \times\:
\left\{ \mathcal{P}_n
\left[-igt\!\cdot\!\d_j\mathcal{A}^-\left(z_{n_2}^+,\hat{\z}_{n_2}\right)\right]
\left[-igt\!\cdot\!\d_j\mathcal{A}^-\left(z_{n_1}^+,\hat{\z}_{n_1}\right)\right]
\prod_{  \begin{array}{c}
           n=1 \\
           n\neq n_1,n_2  \\
         \end{array}
}^{N} \left[-igt\!\cdot\!\mathcal{A}^-\left(z_n^+,\hat{\z}_{n}\right)\right]\right\}
\nn\\
&
+\textrm{NNEik}
\, ,
\label{q_exp_1}
\end{align}
where the first correction comes form the second order Taylor expansion of a $\mathcal{A}^-$ insertion, and the second correction form the first order Taylor expansion of two different $\mathcal{A}^-$ insertions. Rearranging the expression \eqref{q_exp_1} leads to
\begin{align}
{\cal E}  = &
\left\{1 +\frac{i(z_{N}^+\!-\!z_{1}^+)}{8k^+}\, \d_{\z^j} \d_{\z^j}
\right\}
\left\{ \mathcal{P}_n \prod_{n=1}^{N} \left[-igt\!\cdot\!\mathcal{A}^-\left(z_n^+,\hat{\z}_{n}\right)\right]\right\}
\nn\\
&
-\frac{i}{2k^+}\, \sum_{n_1=1}^{N-1}\; \sum_{n_2=n_1+1}^{N} (z_{n_2}^+\!-\!z_{n_1}^+)
\nn\\
& \times\:
\left\{ \mathcal{P}_n
\left[-igt\!\cdot\!\d_j\mathcal{A}^-\left(z_{n_2}^+,\hat{\z}_{n_2}\right)\right]
\left[-igt\!\cdot\!\d_j\mathcal{A}^-\left(z_{n_1}^+,\hat{\z}_{n_1}\right)\right]
\prod_{  \begin{array}{c}
           n=1 \\
           n\neq n_1,n_2  \\
         \end{array}
}^{N} \left[-igt\!\cdot\!\mathcal{A}^-\left(z_n^+,\hat{\z}_{n}\right)\right]\right\}
\nn\\
&
+\textrm{NNEik}
\, .
\label{q_exp_2}
\end{align}
Hence,
\begin{align}
& e^{i(z_{N}^+-z_{1}^+) \frac{(\q-\k)^2}{8k^+}}
\int d^2\z\, e^{-i \z \cdot (\q-\k)}
\;\; {\cal E}
  =
\int d^2\z\, e^{-i \z \cdot (\q-\k)}
\left\{ \mathcal{P}_n \prod_{n=1}^{N} \left[-igt\!\cdot\!\mathcal{A}^-\left(z_n^+,\hat{\z}_{n}\right)\right]\right\}
\nn\\
&
-\frac{i}{2k^+}\int d^2\z\, e^{-i \z \cdot (\q-\k)}\, \sum_{n_1=1}^{N-1}\; \sum_{n_2=n_1+1}^{N} (z_{n_2}^+\!-\!z_{n_1}^+)
\nn\\
& \times\:
\left\{ \mathcal{P}_n
\left[-igt\!\cdot\!\d_j\mathcal{A}^-\left(z_{n_2}^+,\hat{\z}_{n_2}\right)\right]
\left[-igt\!\cdot\!\d_j\mathcal{A}^-\left(z_{n_1}^+,\hat{\z}_{n_1}\right)\right]
\prod_{  \begin{array}{c}
           n=1 \\
           n\neq n_1,n_2  \\
         \end{array}
}^{N} \left[-igt\!\cdot\!\mathcal{A}^-\left(z_n^+,\hat{\z}_{n}\right)\right]\right\}
\nn\\
&
+\textrm{NNEik}
\, ,
\label{q_exp_3}
\end{align}
since the correction obtained after double integration by parts in $\z$ precisely cancels the Next-to-Eikonal correction obtained by expanding the phase factor.

So far, we have performed an analog of what was called the expansion around the classical trajectory in Refs.~\cite{Altinoluk:2014oxa,Altinoluk:2015gia}. It remains to perform what was called the small angle expansion, in which the small parameters are $\hat{\z}_{n}-\z$. In order to reach Next-to-Eikonal accuracy, it is sufficient to expand the leading term in Eq.~\eqref{q_exp_3} to first order in any $\hat{\z}_{n}-\z$, and simply make the replacement $\hat{\z}_{n}\mapsto\z$ in the subleading term in Eq.~\eqref{q_exp_3}. This leads to
\begin{align}
& e^{i(z_{N}^+-z_{1}^+) \frac{(\q-\k)^2}{8k^+}}
\int d^2\z\, e^{-i \z \cdot (\q-\k)}
\;\; {\cal E}
  =
\int d^2\z\, e^{-i \z \cdot (\q-\k)}
\left\{ \mathcal{P}_n \prod_{n=1}^{N} \left[-igt\!\cdot\!\mathcal{A}^-\left(z_n^+,\z\right)\right]\right\}
\nn\\
&
+\frac{(\q^j\!+\!\k^j)}{2k^+}\int d^2\z\, e^{-i \z \cdot (\q-\k)} \sum_{n_1=1}^{N} z_{n_1}^+
\left\{ \mathcal{P}_n
\left[-igt\!\cdot\!\d_j\mathcal{A}^-\left(z_{n_1}^+,\z\right)\right]
\prod_{  \begin{array}{c}
           n=1 \\
           n\neq n_1  \\
         \end{array}
}^{N} \left[-igt\!\cdot\!\mathcal{A}^-\left(z_n^+,\z\right)\right]\right\}
\nn\\
&
-\frac{i}{2k^+}\int d^2\z\, e^{-i \z \cdot (\q-\k)}\, \sum_{n_1=1}^{N-1}\; \sum_{n_2=n_1+1}^{N} (z_{n_2}^+\!-\!z_{n_1}^+)
\nn\\
& \times\:
\left\{ \mathcal{P}_n
\left[-igt\!\cdot\!\d_j\mathcal{A}^-\left(z_{n_2}^+,\z\right)\right]
\left[-igt\!\cdot\!\d_j\mathcal{A}^-\left(z_{n_1}^+,\z\right)\right]
\prod_{  \begin{array}{c}
           n=1 \\
           n\neq n_1,n_2  \\
         \end{array}
}^{N} \left[-igt\!\cdot\!\mathcal{A}^-\left(z_n^+,\z\right)\right]\right\}
\nn\\
&
+\textrm{NNEik}
\, .
\label{q_exp_4}
\end{align}
Inserting this result into Eq.~\eqref{q_prop_pure_A_minus_3}, and using the relations \eqref{Wilson_line} and \eqref{Wilson_line_dagger} to collect field insertions into gauge links, we obtain
\begin{align}
&\delta S_F(x,y) \bigg|_{\textrm{pure }\mathcal{A}^{-}}
=
\int \frac{d^3\uq}{(2\pi)^3} \int \frac{d^3\uk}{(2\pi)^3}\, 2\pi \delta(q^+\!-\!k^+)\,
e^{-i x\cdot \check{q}}\, e^{i y\cdot \check{k}}\, (\slashed{\check{q}}+m)\gamma^+(\slashed{\check{k}}+m)\, \frac{\theta(k^+)}{(2k^+)^2}
\nn\\
& \times\:
\int d^2\z\, e^{-i \z \cdot (\q-\k)}\;
\Bigg\{
\Big[\mathcal{U}_F(x^+,y^+;\z)-\mathbf{1}\Big]
\nn\\
&
+\frac{(\q^j\!+\!\k^j)}{2k^+}\int_{y^+}^{x^+}dv^+\, v^+\, \mathcal{U}_F(x^+,v^+;\z) \left[-igt\!\cdot\!\d_j\mathcal{A}^-\left(v^+,\z\right)\right] \mathcal{U}_F(v^+,y^+;\z)
\nn\\
&
-\frac{i}{2k^+}\int_{y^+}^{x^+}dv^+\,\int_{v^+}^{x^+}dw^+\, (w^+\!-\!v^+)\,
\mathcal{U}_F(x^+,w^+;\z) \left[-igt\!\cdot\!\d_j\mathcal{A}^-\left(w^+,\z\right)\right]
\nn\\
& \hspace{2cm} \times\;\;\;
\mathcal{U}_F(w^+,v^+;\z) \left[-igt\!\cdot\!\d_j\mathcal{A}^-\left(v^+,\z\right)\right] \mathcal{U}_F(v^+,y^+;\z)
\Bigg\}
+\textrm{NNEik}
\label{q_prop_pure_A_minus_4}
\end{align}
for $x^+>y^+$, and
\begin{align}
&\delta S_F(x,y) \bigg|_{\textrm{pure }\mathcal{A}^{-}}
=
\int \frac{d^3\uq}{(2\pi)^3} \int \frac{d^3\uk}{(2\pi)^3}\, 2\pi \delta(q^+\!-\!k^+)\,
e^{-i x\cdot \check{q}}\, e^{i y\cdot \check{k}}\, (\slashed{\check{q}}+m)\gamma^+(\slashed{\check{k}}+m)\, \frac{\theta(-k^+)}{(2k^+)^2}
\nn\\
& \times\:
\int d^2\z\, e^{-i \z \cdot (\q-\k)}\;
\Bigg\{
-\Big[\mathcal{U}_F^{\dag}(y^+,x^+;\z)-\mathbf{1}\Big]
\nn\\
&
-\frac{(\q^j\!+\!\k^j)}{2k^+}\int_{x^+}^{y^+}dv^+\, v^+\, \mathcal{U}_F^{\dag}(v^+,x^+;\z) \left[+igt\!\cdot\!\d_j\mathcal{A}^-\left(v^+,\z\right)\right] \mathcal{U}_F^{\dag}(y^+,v^+;\z)
\nn\\
&
-\frac{i}{2k^+}\int_{x^+}^{y^+}dv^+\,\int_{x^+}^{v^+}dw^+\, (v^+\!-\!w^+)\,
\mathcal{U}_F^{\dag}(w^+,x^+;\z) \left[+igt\!\cdot\!\d_j\mathcal{A}^-\left(w^+,\z\right)\right]
\nn\\
& \hspace{2cm} \times\;\;\;
\mathcal{U}_F^{\dag}(v^+,w^+;\z) \left[+igt\!\cdot\!\d_j\mathcal{A}^-\left(v^+,\z\right)\right] \mathcal{U}_F^{\dag}(y^+,v^+;\z)
\Bigg\}
+\textrm{NNEik}
\label{qbar_prop_pure_A_minus_4}
\end{align}
for $y^+>x^+$.

From the definition of the gauge link \eqref{Wilson_line}, one has the general property
\begin{align}
\d_{\z^j} \mathcal{U}_F\Big(x^+,y^+;\z\Big)
& =
\int_{y^+}^{x^+}dz^+\, \mathcal{U}_F\Big(x^+,z^+;\z\Big)\,
\big[-igt\!\cdot\!\d_{\z^j}\mathcal{A}^-(z^+,\z)\big]\,
\mathcal{U}_F\Big(z^+,y^+;\z\Big)
\label{perp_deriv_of_gauge_link}
\, .
\end{align}
From this relation, and using $(w^+\!-\!v^+)=\int_{v^+}^{w^+} dz^+$, the bilocal operator in Eq.~\eqref{q_prop_pure_A_minus_4} can be written in a compact way as
\begin{align}
& \int_{y^+}^{x^+}dv^+\,\int_{v^+}^{x^+}dw^+\, (w^+\!-\!v^+)\,
\mathcal{U}_F(x^+,w^+;\z) \left[-igt\!\cdot\!\d_j\mathcal{A}^-\left(w^+,\z\right)\right]
\nn\\
& \hspace{2cm} \times\;\;\;
\mathcal{U}_F(w^+,v^+;\z) \left[-igt\!\cdot\!\d_j\mathcal{A}^-\left(v^+,\z\right)\right] \mathcal{U}_F(v^+,y^+;\z)
\nn\\
= &  \int_{y^+}^{x^+} dz^+\, \mathcal{U}_F\Big(x^+,z^+;\z\Big)\,
\overleftarrow{\d_{\z^j}}\, \overrightarrow{\d_{\z^j}}\,
\mathcal{U}_F\Big(z^+,y^+;\z\Big)
\label{q_biloc_op_compact}
\, .
\end{align}
As a remark, the integration over $z^+$ in Eq.~\eqref{q_biloc_op_compact} receives non trivial contributions only from the overlap of the $[y^+, x^+]$ interval with the support $[-L^+/2, L^+/2]$ of the background field.

Similarly, the bilocal operator in Eq.~\eqref{qbar_prop_pure_A_minus_4} can be rewritten as
\begin{align}
& \int_{x^+}^{y^+}dv^+\,\int_{x^+}^{v^+}dw^+\, (v^+\!-\!w^+)\,
\mathcal{U}_F^{\dag}(w^+,x^+;\z) \left[+igt\!\cdot\!\d_j\mathcal{A}^-\left(w^+,\z\right)\right]
\nn\\
& \hspace{2cm} \times\;\;\;
\mathcal{U}_F^{\dag}(v^+,w^+;\z) \left[+igt\!\cdot\!\d_j\mathcal{A}^-\left(v^+,\z\right)\right] \mathcal{U}_F^{\dag}(y^+,v^+;\z)
\nn\\
= &  \int_{x^+}^{y^+} dz^+\, \mathcal{U}_F^{\dag}\Big(z^+,x^+;\z\Big)\,
\overleftarrow{\d_{\z^j}}\, \overrightarrow{\d_{\z^j}}\,
\mathcal{U}_F^{\dag}\Big(y^+,z^+;\z\Big)
\label{qbar_biloc_op_compact}
\, .
\end{align}

The last step is to rewrite the term in the third line of Eq.~\eqref{q_prop_pure_A_minus_4} (or Eq.~\eqref{qbar_prop_pure_A_minus_4}) in a analogous way, in order to get rid of the $v^+$ factor.
For $x^+>y^+$, this can be done thanks to the identity
\begin{align}
v^+ & = \frac{1}{2}\: \left[\int_{z^+_{\min}}^{v^+}dz^+  - \int_{v^+}^{z^+_{\max}}dz^+\right]  +  \frac{(z^+_{\max}\!+\!z^+_{\min})}{2}
\, .
\end{align}
Then,
\begin{align}
& \int d^2\z\, e^{-i \z \cdot (\q-\k)}\; \int_{z^+_{\min}}^{z^+_{\max}}dv^+\, v^+\,
\mathcal{U}_F\Big(z^+_{\max},v^+;\z\Big)\,
\big[-igt\!\cdot\!\d_{\z^j}\mathcal{A}^-(v^+,\z)\big]\,
\mathcal{U}_F\Big(v^+,z^+_{\min};\z\Big)
\nn\\
 =&\, \frac{1}{2}\: \int d^2\z\, e^{-i \z \cdot (\q-\k)}\;
\int_{z^+_{\min}}^{z^+_{\max}}dz^+\, \left[ \d_{\z^j}\mathcal{U}_F\Big(z^+_{\max},z^+;\z\Big)\right]\,
\mathcal{U}_F\Big(z^+,z^+_{\min};\z\Big)
\nn\\
 & -\frac{1}{2}\: \int d^2\z\, e^{-i \z \cdot (\q-\k)}\;
\int_{z^+_{\min}}^{z^+_{\max}}dz^+\, \mathcal{U}_F\Big(z^+_{\max},z^+;\z\Big)\,
\left[ \d_{\z^j}\mathcal{U}_F\Big(z^+,z^+_{\min};\z\Big)\right]
\nn\\
&
+ \frac{(z^+_{\max}\!+\!z^+_{\min})}{2}\, \int d^2\z\, e^{-i \z \cdot (\q-\k)}\; \d_{\z^j}\mathcal{U}_F\Big(z^+_{\max},z^+_{\min};\z\Big)
\nn\\
 =& -\frac{1}{2}\: \int d^2\z\, e^{-i \z \cdot (\q-\k)}\;
\int_{z^+_{\min}}^{z^+_{\max}}dz^+\, \left[ \mathcal{U}_F\Big(z^+_{\max},z^+;\z\Big)\,
\overleftrightarrow{\d_{\z^j}}\,
\mathcal{U}_F\Big(z^+,z^+_{\min};\z\Big)\right]
\nn\\
&
+ \frac{i (z^+_{\max}\!+\!z^+_{\min})}{2}\, (\q^j\!-\!\k^j) \int d^2\z\, e^{-i \z \cdot (\q-\k)}\; \mathcal{U}_F\Big(z^+_{\max},z^+_{\min};\z\Big)
\label{q_loc_op_compact}
\, .
\end{align}
In general, $z^+_{\min}$ and $z^+_{\max}$ are the bounds of the interval formed by the intersection of $[y^+, x^+]$ and $[-L^+/2, L^+/2]$.

In the particular case of the quark propagation through the whole medium, meaning $x^+> L^+/2$ and $y^+<-L^+/2$, we have $z^+_{\max}=L^+/2$ and $z^+_{\min}=-L^+/2$, so that $z^+_{\max}+z^+_{\min}=0$, which implies that the extra term proportional to the eikonal contribution in Eq.~\eqref{q_loc_op_compact} vanishes. All in all, we obtain the result \eqref{q_prop_ba_pure_A_minus}.

Otherwise, if either $x^+$ or $y^+$ lies within the medium, the interval $[z^+_{\min}, z^+_{\max}]$ is in general not centered around $0$, leading to a non-vanishing of the extra term in Eq.~\eqref{q_loc_op_compact}. It can however be resummed into a phase factor multiplying the eikonal contribution, since a translation in $z^+$ is equivalent to a translation in $\z$, via Eqs.~\eqref{zn_to_un_change_of_var} and \eqref{un_to_wn_and_z_change_of_var}.

For $x^+<y^+$, the situation is analog. In particular, for $y^+> L^+/2$ and $x^+<-L^+/2$, one finds the result \eqref{qbar_prop_ba_pure_A_minus}.
\section{Comparison with the earlier results in the literature}
\label{Sec:Comparison}

The NEik corrections to the quark propagator in a gluon background have been also calculated in Ref. \cite{Chirilli:2018kkw}. It has been observed recently in Ref. \cite{Chirilli:2021lif} that the results of Ref. \cite{Chirilli:2018kkw} and of the present manuscript 
show some apparent differences. The purpose of this appendix is to discuss such possible disagreements. 

In Ref. \cite{Chirilli:2018kkw}, the NEik corrections to the quark propagator are written in terms two of  different operators $\hat {\cal O}_1$ and $\hat {\cal O}_2$.  The operator $\hat {\cal O}_1$ contains both a helicity dependent part (which agrees with our result) and an unpolarized part which seems different from our unpolarized contribution given in Eq. \eqref{q_prop_ba_h_indep}. The operator $\hat {\cal O}_2$ contain new structures which are not present in our result. Moreover, in Ref.\cite{Chirilli:2021lif}, this operator $\hat {\cal O}_2$ is further split into two pieces $\hat {\cal O}_j$ and $\hat {\cal O}_{\bullet *}$. 
Let us discuss each contribution separately.
\begin{enumerate}

\item{Contribution $\hat {\cal O}_1$:} 

The contribution $\hat {\cal O}_1$ is defined in Eq. (4.13) in Ref. \cite{Chirilli:2018kkw} and it contains three terms. The first term which includes a ${\cal F}_{ij}$ insertion agrees with our helicity dependent contributions to the quark and antiquark propagators defined in Eqs. \eqref{q_prop_ba_h_dep} and \eqref{qbar_prop_ba_h_dep}. 

The last term in Eq. (4.13) in Ref. \cite{Chirilli:2018kkw} is consistent with the last term in Eq. \eqref{q_prop_ba_h_indep} (or in Eq. \eqref{qbar_prop_ba_h_indep} for the antiquark) in the present manuscript. This can be shown as follows. In the most general case, we have the identity: 
\begin{eqnarray}
\partial_{\z^i}{\cal U}_F(x^+,y^+;\z)&=& -igt\cdot {\cal A}_i(x^+,\z) {\cal U}_F(x^+,y^+;\z)+ig\; {\cal U}_F(x^+,y^+;\z) t\cdot {\cal A}_i(y^+,\z) 
\nonumber\\
&+&ig\int_{y^+}^{x^+} dz^+\;  {\cal U}_F(x^+,z^+;\z) \,\big[ t\cdot {{\cal F}^{-}}_i(z^+,\z) \big]\; {\cal U}_F(z^+,y^+;\z)
\label{identity_wrt_G}
\end{eqnarray}
Note that this identity is provided in Eqs. (A.2) and (A.3) in Ref. \cite{Chirilli:2018kkw}.\footnote{Note that a different convention for the definition of the covariant derivative is adopted in our work and in Ref. \cite{Chirilli:2018kkw}. The translation between the two conventions can be performed by replacing the coupling $g$ by $-g$ everywhere.}  Thanks to this identity, the last line of Eq. \eqref{q_prop_ba_h_indep} can be written as 
\begin{eqnarray}
&&
\hspace{-1cm}
-\frac{i}{2k^+}\int_{-\frac{L^+}{2}}^{\frac{L^+}{2}}dz^+\,  \left[\mathcal{U}_F\Big(\frac{L^+}{2},z^+;\z\Big)\,
\overleftarrow{\mathcal{D}_{\z^j}}\, \overrightarrow{\mathcal{D}_{\z^j}}\,
\mathcal{U}_F\Big(z^+,-\frac{L^+}{2};\z\Big)\right]\nonumber\\
&&=
-\frac{i}{2k^+}\int_{-\frac{L^+}{2}}^{\frac{L^+}{2}}dz^+ \int_{z^+}^{\frac{L^+}{2}} d\omega'^{+}\int^{z^+}_{-\frac{L^+}{2}} d\omega^{+} 
{\cal U}_F\bigg(\frac{L^+}{2},\omega'^+;\z\bigg) \big[ig\,  t\cdot {{\cal F}^{-}}_j(\omega'^+,\z) \big] \nonumber\\
&&
\hspace{1cm}
\times\;  
{\cal U}_F(\omega'^+,\omega^+;\z)  \big[ig\,  t\cdot {{\cal F}^{-}}_j(\omega^+,\z) \big] {\cal U}_F\bigg(\omega^+,-\frac{L^+}{2}^+;\z\bigg) \\
&&=
-\frac{i}{2k^+}\int_{-\frac{L^+}{2}}^{\frac{L^+}{2}}d\omega^+ 
\int_{\omega^+}^{\frac{L^+}{2}} d\omega'^{+}\; 
(\omega^+-\omega'^+) \; 
{\cal U}_F\bigg(\frac{L^+}{2},\omega'^+;\z\bigg) \big[ig\,  t\cdot {{\cal F}^{-}}_j(\omega'^+,\z) \big] \nonumber\\
&&
\hspace{1cm}
\times\;  
{\cal U}_F(\omega'^+,\omega^+;\z)  \big[ig\,  t\cdot {{\cal F}^{-}}_j(\omega^+,\z) \big] {\cal U}_F\bigg(\omega^+,-\frac{L^+}{2}^+;\z\bigg)
\label{b3}
\end{eqnarray}
where we have used the assumption that the gauge field vanishes both outside and at the edge of the medium.  Then, Eq. \eqref{b3} is equivalent to the last line of Eq. (4.13) in Ref. \cite{Chirilli:2018kkw} within this assumption. 

Let us now compare the second term in $\hat {\cal O}_1$ given in Eq. (4.13) of Ref. \cite{Chirilli:2018kkw} with the second term of our Eq. \eqref{q_prop_ba_h_indep}. Again, by using the identity given in Eq. \eqref{identity_wrt_G}, the second term in our Eq. \eqref{q_prop_ba_h_indep} can be written as 
\begin{eqnarray}
&&
-\frac{(\q^j\!+\!\k^j)}{4k^+}\int_{-\frac{L^+}{2}}^{\frac{L^+}{2}}dz^+\, \left[\mathcal{U}_F\Big(\frac{L^+}{2},z^+;\z\Big)\,
\overleftrightarrow{\mathcal{D}_{\z^j}}\,
\mathcal{U}_F\Big(z^+,-\frac{L^+}{2};\z\Big)\right]\nonumber\\
&&
= -\frac{(\q^j\!+\!\k^j)}{4k^+}\int_{-\frac{L^+}{2}}^{\frac{L^+}{2}}dz^+\, \int_{-\frac{L^+}{2}}^{z^+} d\omega^+ 
{\cal U}_F\bigg(\frac{L^+}{2},\omega^+;\z\bigg) \big[ig\,  t\cdot {{\cal F}^{-}}_j(\omega^+,\z) \big] {\cal U}_F\bigg(\omega^+, -\frac{L^+}{2};\z\bigg) \nonumber\\
&&
 +\frac{(\q^j\!+\!\k^j)}{4k^+}\int_{-\frac{L^+}{2}}^{\frac{L^+}{2}}dz^+\, \int_{z^+}^{\frac{L^+}{2}} d\omega^+ 
{\cal U}_F\bigg(\frac{L^+}{2},\omega^+;\z\bigg) \big[ig\,  t\cdot {{\cal F}^{-}}_j(\omega^+,\z) \big] {\cal U}_F\bigg(\omega^+, -\frac{L^+}{2};\z\bigg) \nonumber\\
&&
= -\frac{(\q^j\!+\!\k^j)}{4k^+}\int_{-\frac{L^+}{2}}^{\frac{L^+}{2}}d\omega^+\bigg\{ \bigg[\frac{L^+}{2}-\omega^+\bigg]-\bigg[\omega^+-\bigg(-\frac{L^+}{2}\bigg)\bigg]\bigg\}\nonumber\\
&&
\hspace{3.5cm}
\times\; 
{\cal U}_F\bigg(\frac{L^+}{2},\omega^+;\z\bigg) \big[ig\,  t\cdot {{\cal F}^{-}}_j(\omega^+,\z) \big] {\cal U}_F\bigg(\omega^+, -\frac{L^+}{2};\z\bigg)\nonumber\\
&&
= \frac{(\q^j\!+\!\k^j)}{2k^+} \int_{-\frac{L^+}{2}}^{\frac{L^+}{2}}d\omega^+ \, \omega^+ \, {\cal U}_F\bigg(\frac{L^+}{2},\omega^+;\z\bigg) \big[ig\,  t\cdot {{\cal F}^{-}}_j(\omega^+,\z) \big] {\cal U}_F\bigg(\omega^+, -\frac{L^+}{2};\z\bigg)
\end{eqnarray}
which is equivalent to second term in $\hat {\cal O}_1$ given in Eq. (4.13) of Ref. \cite{Chirilli:2018kkw}, since the action of the anticommutator with the ${\hat p}^j$ operator gives the sum of the incoming and outgoing transverse momenta, ($\q^j+\k^j$) in our notation. 

So far, we have shown a full agreement between our complete result and the $\hat {\cal O}_1$ contribution in Ref. \cite{Chirilli:2018kkw},  whereas contributions of the type $\hat {\cal O}_{\bullet *}$ and $\hat {\cal O}_j$ do not arise in our computation.  
\item{Contribution $\hat {\cal O}_{\bullet *}$:}

The contribution $\hat {\cal O}_{\bullet *}$ corresponds to a Wilson line decorated by an insertion of the field strength ${\cal F}_{+-}$, and can be written in our notations as

\begin{equation}
\hat {\cal O}_{\bullet *}\sim \frac{1}{k^+}\int_{y^+}^{x^+} dz^+\, {\cal U}_F(x^+,z^+;\z) \, g \, t\cdot {\cal F}_{+-} (\uz)\,  {\cal U}_F(z^+,y^+;\z) 
\label{O_dot_star}
\end{equation}

By definition, one has in general
\begin{equation}
t \cdot {\cal F}_{+-}(z)\equiv t\cdot \big\{\partial_+ {\cal A}^+(z)- \partial_{-}{\cal A}^-(z) \big\}+ ig\big[t\cdot {\cal A}^-(z), t\cdot {\cal A}^+(z) \big]
\label{F_plus_minus}
\end{equation}

Both in the present manuscript and in Refs. \cite{Chirilli:2018kkw,Chirilli:2021lif}, the background field is assumed to be $z^-$ independent. In this case Eq. \eqref{F_plus_minus} reduces to 
\begin{equation}
t \cdot {\cal F}_{+-}(\uz)= t\cdot \partial_+ {\cal A}^+(\uz)+ ig\big[t\cdot {\cal A}^-(\uz), t\cdot {\cal A}^+(\uz) \big]=\Big[ {\cal D}_{z^+}\, ,\,  t\cdot \!{\cal A}^+(\uz) \Big]
\label{F_plus_minus_1}
\end{equation}

Inserting Eq. \eqref{F_plus_minus_1} into Eq. \eqref{O_dot_star} and performing an integration by parts in one of the terms, we get
\begin{eqnarray}
\hat {\cal O}_{\bullet *}\sim && \frac{1}{k^+}\int dz^+\, \theta(x^+-z^+) \, {\cal U}_F(x^+,z^+;\z) \, \Big\lgroup - \overleftarrow{{\cal D}_{z^+}} \; g t\cdot {\cal A}^{+} (\uz)\,  -   g t\cdot {\cal A}^{+} (\uz)\; \overrightarrow{{\cal D}_{z^+}}  \Big\rgroup \nonumber\\
&&
\hspace{4cm}
\times \;  {\cal U}_F(z^+,y^+;\z) \,  \theta(z^+-y^+) \nonumber\\
\sim
&&
\frac{1}{k^+}\bigg\lgroup g t\cdot {\cal A}^{+} (x^+,\z)\; {\cal U}_F(x^+,y^+;\z)- {\cal U}_F(x^+,y^+;\z)\; g t\cdot {\cal A}^{+} (y^+,\z)\bigg\rgroup
\label{O_dot_star_final}
\end{eqnarray}
using the vanishing of the action of the covariant derivative on the gauge links, leaving only ordinary derivative acting on the $\theta$-functions. 

In our setup,  the gauge fields are taken to be vanishing outside of the medium. 
Then, Eq. \eqref{O_dot_star_final} implies that the $\hat {\cal O}_{\bullet *}$ contribution vanishes when both ends of the propagator are outside of the medium. This observation is consistent with the fact that Eqs. \eqref{q_prop_ba_h_indep} and \eqref{q_prop_ba_h_dep} do not include a contribution of the type $\hat {\cal O}_{\bullet *}$ at NEik accuracy. If we allow the gauge field to be non-vanishing outside of the medium, this contribution would be still of NNEik order due to the insertion of the suppressed ${\cal A}^{+}$ component. This discussion is consistent with Appendix C in Ref. \cite{Chirilli:2018kkw}. 

\item{Contribution $\hat {\cal O}_j$:}

 Concerning the contribution $\hat {\cal O}_j$, the situation is more complicated. It requires an in-depth study of the derivation performed in \cite{Chirilli:2018kkw} which is in progress \cite{Altinoluk_Beuf}. Here, we simply outline and discuss preliminary results of that  study. 

A major difference between the methods employed in this work and in Ref. \cite{Chirilli:2018kkw} to derive the NEik corrections to the quark propagator is the treatment of the symmetry between the end points $x$ and $y$. In our derivation based on Feynman diagrams, this symmetry is maintained at each step of the calculation. By contrast, in the derivation performed in Ref. \cite{Chirilli:2018kkw}, an asymmetry between the end points $x$ and $y$ is introduced at the very first step by putting the operator $\hat{\slashed{P}}$ on the left in the numerator (see Eq. 4.1 in Ref. \cite{Chirilli:2018kkw}). This is equivalent to focusing on the equation of motion in $x$ for the quark propagator. The other possible choice would be to put the operator $\hat{\slashed{P}}$ on the right in the numerator or equivalently to consider the equation of motion in $y$. Both choices have to provide the same result eventually, even though intermediate steps differ. 
In Ref. \cite{Chirilli:2018kkw}, the derivation is initially performed with the first choice only, and an asymmetric result in $x$ and $y$ is obtained (see Eq. 4.17). This asymmetry is carried by the term containing the operator ${\cal \hat O}_2$, which is split into the operators  $\hat {\cal O}_j$ and  $\hat {\cal O}_{\bullet *}$ in Ref. \cite{Chirilli:2021lif}. In Ref. \cite{Chirilli:2018kkw}, an additional step is then introduced to restore the symmetry with respect to $x$ and $y$ (see Eq. 4.18) by taking the average of the results obtained from the two choices (putting the operator $\hat{\slashed{P}}$ either on the left or on the right in the numerator), even though both results should be equivalent. 

For a better analysis of the situation, it is convenient to introduce the projections ${\cal P}_G\equiv \gamma^-\gamma^+/2$ and ${\cal P}_B\equiv\gamma^+\gamma^-/2$ on the so-called good and bad components of Dirac spinors. When we apply the good or the bad projections on each side of the equation of motion with respect to either $x$ or $y$ for the quark propagator, one ends up with coupled equations for four blocks in the quark propagator: ${\cal P}_GS_F(x,y){\cal P}_B$,  ${\cal P}_GS_F(x,y){\cal P}_G$,  ${\cal P}_BS_F(x,y){\cal P}_B$ and  ${\cal P}_BS_F(x,y){\cal P}_G$. These equations can be solved order by order as an eikonal expansion  using Green's function methods.

Our preliminary results from Ref. \cite{Altinoluk_Beuf} indicates that the ${\cal P}_GS_F(x,y){\cal P}_B$ block does not receive corrections of the type $\hat {\cal O}_j$ which is in agreement with the results of Refs. \cite{Chirilli:2018kkw} and \cite{Chirilli:2021lif} after projection. A more illustrative example is provided by the ${\cal P}_GS_F(x,y){\cal P}_G$ block. From the equation of motion with respect to $y$, we obtain the following expression 
\begin{eqnarray}
\hspace{-1cm}
{\cal P}_GS_F(x,y){\cal P}_G&=&\bigg\{G_F(x,y)+\int d^Dz\, G_F(x,y)
\bigg\lgroup \frac{[\gamma^i, \gamma^j]}{4}gt\cdot\!{\cal F}_{ij}(z)
\label{good_good_right}\\
&&
\hspace{2.2cm}
%
- gt\cdot\!{\cal F}_{+-}(z)\bigg\rgroup 
G_F(z,y)\bigg\} {\cal P}_G \Big[- i\gamma^l\overleftarrow{\cal D}_{y^l}+m\Big]+{\rm NNEik}\nonumber
\end{eqnarray}
where $G_F(x,y)$ is the full scalar propagator in gluon background, which includes the subeikonal corrections associated with the transverse motion within the target. This equation is valid for arbitrary $x^-$ dependence of the background field, but $y$ is assumed to be outside of the medium. Note that, this expression involves contributions of the type $\hat {\cal O}_1$ and  $\hat {\cal O}_{\bullet *}$ but not of the type $\hat {\cal O}_j$. If one considers the equation of motion with respect to $x$ for the ${\cal P}_GS_F(x,y){\cal P}_G$ block, one obtains a much more complicated expression which at the intermediate steps include contributions apparently related to $\hat {\cal O}_j$. Nevertheless, the equation of motion for $x$ and $y$ should give equivalent results. In Ref. \cite{Chirilli:2018kkw}, the equation of motion for $x$ is considered and the result includes contribution of the type $\hat {\cal O}_j$. However, Eq. \eqref{good_good_right} shows that  ${\cal P}_GS_F(x,y){\cal P}_G$ block can be written without $\hat {\cal O}_j$ type of contribution. 

For the ${\cal P}_BS_F(x,y){\cal P}_B$ block, the situation is reversed. The equation of motion with respect to  $x$ is quite simple and reads 
\begin{eqnarray}
\hspace{-1cm}
{\cal P}_BS_F(x,y){\cal P}_B&=& {\cal P}_B \Big[i\gamma^l\overrightarrow{\cal D}_{x^l}+m\Big] \bigg\{G_F(x,y)+\int d^Dz\, G_F(x,y)
\bigg\lgroup \frac{[\gamma^i, \gamma^j]}{4}gt\cdot\!{\cal F}_{ij}(z)
\nonumber\\
&&
\hspace{4.5cm}
%
+ gt\cdot\!{\cal F}_{+-}(z)\bigg\rgroup 
G_F(z,y)\bigg\} +{\rm NNEik}
\label{bad_bad_left}
\end{eqnarray}
It does not involve contributions of the type $\hat {\cal O}_j$. This is in agreement with the result of Ref. \cite{Chirilli:2018kkw} after projection.  On the other hand, the equation of motion with respect to $y$ in this block is complicated, despite the fact that it should be equivalent to \eqref{bad_bad_left}. Therefore, a convenient strategy in order to obtain a result with an overall symmetry with respect to $x$ and $y$ is to use Eq. \eqref{good_good_right} for the ${\cal P}_GS_F(x,y){\cal P}_G$ block and Eq. \eqref{bad_bad_left} for the ${\cal P}_BS_F(x,y){\cal P}_B$ block. In this way, one can avoid the appearance of $\hat {\cal O}_j$ type contributions both in the ${\cal P}_GS_F(x,y){\cal P}_G$ and ${\cal P}_BS_F(x,y){\cal P}_B$ blocks. By contrast, 
the symmetrization procedure performed in Eq. 4.18 in Ref. \cite{Chirilli:2018kkw} produces an $\hat {\cal O}_j$ type contribution in both of these blocks.


Finally, both of the equations of motion with respect to $x$ and $y$ for the remaining ${\cal P}_BS_F(x,y){\cal P}_G$ block are more complicated. A detailed study to understand  if $\hat {\cal O}_j$ type of contributions are avoidable or not for this block is in progress \cite{Altinoluk_Beuf}. 

%


\end{enumerate}
 

\section{Forward antiquark-nucleus scattering at NEik accuracy}
\label{Sec:antiquark-observables}

In this appendix, we discuss the use of the antiquark propagator at next-to-eikonal accuracy given in Eq. \eqref{unpol_hdep_decomp} (with the unpolarized piece given in Eq. \eqref{qbar_prop_ba_h_indep} and the helicity dependent piece given in Eq. \eqref{qbar_prop_ba_h_dep}) for  the computation of the forward antiquark scattering on a dense target. Similar to the analysis performed for quarks presented in Section \ref{sec:cross_section}, here we consider both the unpolarized antiquark cross section and antiquark helicity asymmetry at the partonic level. 


Starting from the definition of the free fermion fields given in Eqs. \eqref{Psi-1} and \eqref{Psi-2} and using the fact that spinors satisfy
\ba
\label{eq:condition-spinors-v}
\bar v(\check{p},h) \gamma^+ v(\check{k},h') = \sqrt{2p^+}\sqrt{2k^+} \delta_{hh'}\, ,
\ea
one can easily write the annihilation and creation operators of an antiquark in terms of the free fermion fields as
\ba
\label{d_op}
\hat d(\check{k},h,\alpha) &=& \int d^2 \x \int dx^- e^{ix\cdot \check{k}}
\bar \Psi_\alpha(x) \gamma^+ v(\check{k},h),
\\
\label{d_dag_op}
\hat d^\dagger(\check{k},h,\alpha) &=& \int d^2 \x \int dx^- e^{-ix\cdot \check{k}}
\bar v(\check{k},h) \gamma^+ \Psi_\alpha(x).
\ea

Similar to the case of quarks, in order to construct the $S$-matrix one needs to distinguish between the incoming and outgoing free antiquark fields. Therefore, we write the antiquark creation and annihilation operators as
\ba
\label{d_op_out}
\hat d_{\rm out}(\check{p}_2,h,\alpha) &=& \lim_{y^+\to +\infty} \int d^2 \y \int dy^- e^{iy\cdot \check{p}_2}
\bar \Psi_\alpha(y) \gamma^+ v(\check{p}_2,h),
\\
\label{d_dag_op_in}
\hat d_{\rm in}^\dagger(\check{p}_1,h',\beta) &=& \lim_{x^+\to -\infty}\int d^2 \x \int dx^- e^{-ix\cdot \check{p}_1}
\bar v(\check{p}_1,h') \gamma^+ \Psi_\beta(x).
\ea

The formal definition of the $S$-matrix in terms of the incoming and outgoing antiquark operators is given by
\ba
\label{eq:S-matrix-def-qbar}
S_{\bar{q}(\check{p}_2,h,\alpha)\leftarrow \bar{q}(\check{p}_1,h',\beta) }  &=& \langle 0| \hat d_{\rm out} (\check{p}_2,h,\alpha)
\hat d^\dagger_{\rm in} (\check{p}_1,h',\beta) | 0 \rangle\, .
\ea
It is now straightforward to write the $S$-matrix as
\ba
\label{eq:S-matrix-qbar}
S_{\bar{q}(\check{p}_2,h,\alpha)\leftarrow \bar{q}(\check{p}_1,h',\beta) } &=& \lim_{y^+\to +\infty} \lim_{x^+ \to -\infty} \int d^2 \x \int dx^- \! \! \int d^2 \y \int dy^-
e^{-ix\cdot {\check{p}_1} + iy\cdot {\check{p}_2} }  \nn \\
&& \qquad\qquad \times
\langle 0|\hat T  \left[\bar \Psi_\alpha(y) \gamma^+ v(\check{p}_2,h)\right]
\left[\bar v(\check{p}_1,h') \gamma^+ \Psi_\beta(x)\right]
| 0 \rangle
\nn \\
&=& \lim_{y^+\to +\infty} \lim_{x^+ \to -\infty} \int d^2 \x \int dx^- \! \! \int d^2 \y \int dy^-
e^{-ix\cdot {\check{p}_1} + iy\cdot {\check{p}_2} }
\nn \\
&& \qquad\qquad \times (-1)\,
 \bar v(\check{p}_1,h') \gamma^+ S_F(x,y)_{\beta \alpha}
\gamma^+ v(\check{p}_2,h)\, ,
\ea
where we have used the definition of the Feynman propagator given in Eq. \eqref{eq:propagator} in the second equality. The factor of $(-1)$ in the second equality originates from anticommuting the free fermion fields in order to get the form of the Feynman propagator defined in Eq. \eqref{eq:propagator}.

After introducing the antiquark scattering amplitude ${\overline{\cal M}}^{hh'}_{\alpha\beta}(\p_2,\underline{p_1})$, defined from the $S$-matrix through
\be
S_{\bar{q}(\check{p}_2,h,\alpha)\leftarrow \bar{q}(\check{p}_1,h',\beta) }=(2p_1^+)2\pi\, \delta(p_2^+-p_1^+)\, i{\overline{\cal M}}^{hh'}_{\alpha\beta}(\p_2,\underline{p_1})
\ee
and using the Feynman propagator for the antiquark at next-to-eikonal accuracy given in Eq. \eqref{unpol_hdep_decomp} (together with Eqs. \eqref{qbar_prop_ba_h_indep} and \eqref{qbar_prop_ba_h_dep}) one gets the following expression for the antiquark scattering amplitude:
\ba
\label{eq:amplitude-spinors-antiquark}
i {\overline{\cal M}}^{hh'}_{\alpha\beta}(\p_2,\underline{p_1})&=&
\frac{1}{2p_1^+}
\int d^2 \z\, e^{+i\z\cdot(\p_1\!-\!\p_2)} \, \bar v(\check{p_1},h') \gamma^+
\Bigg\{\mathcal{U}^\dagger_F\Big(\frac{L^+}{2},-\frac{L^+}{2};\z\Big)
\\
&&
\hspace{-0.5cm}
+\frac{i[\gamma^i,\gamma^j]}{8p_2^+}\int_{-\frac{L^+}{2}}^{\frac{L^+}{2}}dz^+\,
\left[\mathcal{U}^\dagger_F\Big(z^+,-\frac{L^+}{2};\z\Big)\,
\Big(-igt \!\cdot \!\mathcal{F}_{ij}(\uz) \Big)
\mathcal{U}^\dagger_F\Big(\frac{L^+}{2},z^+;\z\Big)\right]
\nn\\
&&
\hspace{-0.5cm}
-\frac{i}{2p_1^+}\int_{-\frac{L^+}{2}}^{\frac{L^+}{2}}dz^+\,  \left[\mathcal{U}^\dagger_F\Big(z^+, -\frac{L^+}{2};\z\Big)\,
\overleftarrow{\mathcal{D}_{\z^j}}\, \overrightarrow{\mathcal{D}_{\z^j}}\,
\mathcal{U}^\dagger_F\Big(\frac{L^+}{2},z^+;\z\Big)\right]
\nn\\
&&
\hspace{-0.5cm}
+\frac{(\p_1^j\!+\!\p_2^j)}{4p_1^+}\int_{-\frac{L^+}{2}}^{\frac{L^+}{2}}dz^+\, \left[\mathcal{U}^\dagger_F\Big(z^+,-\frac{L^+}{2};\z\Big)\,
\overleftrightarrow{\mathcal{D}_{\z^j}}\,
\mathcal{U}^\dagger_F\Big(\frac{L^+}{2},z^+;\z\Big)\right]
\Bigg\}_{\alpha\beta} v(\check{p_2},h) \, ,\nn
\ea
where we have used Eqs. \eqref{Spinor_algebra_1} and \eqref{Spinor_algebra_2} to simplify the Dirac structure.  Moreover, again realizing that $[\gamma^i,\gamma^j]=-4i\epsilon^{ij} S^3$ and using the action of helicity operator $S^3$ on the spinor  $v(\check{k},h)$ which is given in Eq. \eqref{S3-on-v}, one can immediately realize that
\be
\bar v(\check{p}_1,h')\gamma^+\frac{[\gamma^i,\gamma^j]}{4}v(\check{p}_2,h)=i\epsilon^{ij}h\bar v(\check{p}_1,h')\gamma^+v(\check{p}_2,h) \, ,
\ee
 which can be used to further simplify the antiquark scattering amplitude into
 \ba
\label{eq:amplitude-spinors-antiquark-2}
i {\overline{\cal M}}^{hh'}_{\alpha\beta}(\p_2,\underline{p_1})&=& \delta_{hh'}
\int d^2 \z\, e^{+i\z\cdot(\p_1\!-\!\p_2)} \,
\Bigg\{\mathcal{U}^\dagger_F\Big(\frac{L^+}{2},-\frac{L^+}{2};\z\Big)
\\
&&
\hspace{-0.5cm}
-\frac{\epsilon^{ij}h}{2p_1^+}
\int_{-\frac{L^+}{2}}^{\frac{L^+}{2}}dz^+\,
\left[\mathcal{U}^\dagger_F\Big(z^+,-\frac{L^+}{2};\z\Big)\,
\Big(-igt \!\cdot \!\mathcal{F}_{ij}(\uz) \Big)
\mathcal{U}^\dagger_F\Big(\frac{L^+}{2},z^+;\z\Big)\right]
\nn\\
&&
\hspace{-0.5cm}
-\frac{i}{2p_1^+}\int_{-\frac{L^+}{2}}^{\frac{L^+}{2}}dz^+\,  \left[\mathcal{U}^\dagger_F\Big(z^+, -\frac{L^+}{2};\z\Big)\,
\overleftarrow{\mathcal{D}_{\z^j}}\, \overrightarrow{\mathcal{D}_{\z^j}}\,
\mathcal{U}^\dagger_F\Big(\frac{L^+}{2},z^+;\z\Big)\right]
\nn\\
&&
\hspace{-0.5cm}
+\frac{(\p_1^j\!+\!\p_2^j)}{4p_1^+}\int_{-\frac{L^+}{2}}^{\frac{L^+}{2}}dz^+\, \left[\mathcal{U}^\dagger_F\Big(z^+,-\frac{L^+}{2};\z\Big)\,
\overleftrightarrow{\mathcal{D}_{\z^j}}\,
\mathcal{U}^\dagger_F\Big(\frac{L^+}{2},z^+;\z\Big)\right]
\Bigg\}_{\alpha\beta} \, .\nn
\ea
This expression is the final form of the antiquark scattering amplitude at NEik accuracy.

Our next order of business is to use the antiquark scattering amplitude to compute the unpolarized antiquark cross section at partonic level. When written in terms of the antiquark scattering amplitude, it reads
\ba
\label{eq:dif-cross-section-antiquark}
\frac{d^2\sigma^{\bar qA\to \bar q+X}}{d^2\p_2} 
&=& \frac{1}{2N_c} \frac{1}{(2\pi)^2}\sum_{h,h'} \sum_{\alpha,\beta}
{\overline{\cal M}}^{hh'}_{\alpha\beta}(\p_2,\underline{p_1})^\dagger {\overline{\cal M}}^{hh'}_{\alpha\beta}(\p_2,\underline{p_1}) \bigg|_{ p_2^+=p_1^+}
\ea
after performing the integration over $p_2^+$.
After inserting Eq. \eqref{eq:amplitude-spinors-antiquark-2} into Eq. \eqref{eq:dif-cross-section-antiquark}, one arrives to the unpolarized antiquark partonic level cross section which reads
\ba
\label{eq:cross-section-unpol-explicit-antiquark}
\frac{d^2\sigma^{\bar qA\to \bar q+X}}{d^2\p_2}
&=& \frac{1}{N_c} \frac{1}{(2\pi)^2}
\int d^2 \z' \int d^2 \z\,  e^{-i(\p_2-\p_1)\cdot(\z\!-\!\z')} \,
{\rm tr}
\Bigg\{\mathcal{U}_F(\z') \mathcal{U}^\dagger _F(\z) \\
&&
\hspace{-1.2cm}
+\frac{1}{2p_1^+}\int_{-\frac{L^+}{2}}^{\frac{L^+}{2}}dz^+\,
\mathcal{U}_F(\z' )\,
\mathcal{U}_F ^\dagger\Big(z^+,-\frac{L^+}{2};\z\Big)\,
\left[ -i\overleftarrow{\mathcal{D}_{\z^j}}\, \overrightarrow{\mathcal{D}_{\z^j}}\,
+\frac{(\p_1^j+\p_2^j)}{2} \overleftrightarrow{\mathcal{D}_{\z^j}} \right]
\mathcal{U}^\dagger_F\Big(\frac{L^+}{2},z^+;\z\Big)
\nn\\
&&
\hspace{-1.2cm}
+\frac{1}{2p_1^+}\int_{-\frac{L^+}{2}}^{\frac{L^+}{2}}dz^+\,
\mathcal{U}_F \Big(\frac{L^+}{2},z^+;\z'\Big)\,
\left[ i\overleftarrow{\mathcal{D}_{\z'^j}}\, \overrightarrow{\mathcal{D}_{\z'^j}}\,
-\frac{(\p_1^j+\p_2^j)}{2} \overleftrightarrow{\mathcal{D}_{\z'^j}} \right]
\mathcal{U}_F \Big(z^+,-\frac{L^+}{2};\z'\Big)
\mathcal{U}^\dagger_F (\z)
\Bigg\} \, .\nn
\ea
We can now introduce the target averaging process. Changing the integration variables to $ \z-\z'\equiv\r$ and $\z+\z' \equiv 2\b$ and using the same definitions for the fundamental dipole and decorated dipole operators after integration over $\b$ (given in Eqs. \eqref{dipole}, \eqref{decorated-dipole-1} and \eqref{decorated-dipole-2} respectively), the target-averaged unpolarized cross section for an antiquark can be written as
\ba
\hspace{-0.3cm}
\bigg\langle \frac{d^2\sigma^{\bar qA\to \bar q+X}}{d^2\p_2} \bigg\rangle_A&=&\frac{1}{(2\pi)^2}\int d^2\r \; e^{-i(\p_2-\p_1)\cdot\r}\bigg\{ d^\dagger_F(\r)-\frac{1}{2p_1^+}\frac{(\p_1^j+\p_2^j)}{2}\Big[ {\cal O}^{\dagger j}_{(1)}(\r)+{\cal O}_{(1)}^j(-\r)\Big]
 \nonumber\\
&&
\hspace{5.5cm}
-\frac{i}{2p_1^+}\Big[ {\cal O}^{\dagger}_{(2)}(\r)-{\cal O}_{(2)}(-\r)\Big]\bigg\} \, .
\ea
After using the symmetry properties of the decorated dipoles given in Eqs. \eqref{O1_sym} and \eqref{O2_sym}, and using the definitions of the Pomeron and Odderon operators introduced in Eqs. \eqref{bar-Pomeron} and \eqref{bar-Odderon}, one can easily arrive to Eq. \eqref{final-antiquark-unpolarized} for the target-averaged unpolarized cross section of an antiquark. Note that, in order to be able to compare with Eq. \eqref{final-quark-unpolarized}, we have renamed $\p_2\mapsto\q$ and $\underline{p_1}\mapsto \underline{k}$ in the final result when we are writing Eq. \eqref{final-antiquark-unpolarized}.

Let us finally perform the computation of the antiquark helicity asymmetry. As explained for the quark case, here we consider just the numerator of the antiquark helicity asymmetry and define it as
\ba
\frac{d^2\Delta\sigma^{ \bar{q}A\to  \bar{q}+X}}{d^2\p_2}\equiv \frac{d^2\sigma^{ \bar{q}^{(+)}A\to  \bar{q}+X}}{d^2\p_2} -\frac{d^2\sigma^{ \bar{q}^{(-)}A\to  \bar{q}+X}}{d^2\p_2}
\ea
which can be written in terms of the antiquark scattering amplitude as
\ba
\label{eq:dif-cross-section-pol-antiquark}
\frac{d^2\Delta\sigma^{ \bar{q}A\to  \bar{q}+X}}{d^2\p_2}=\frac{1}{2N_c} \frac{1}{(2\pi)^2}\sum_{h,h'} \sum_{\alpha,\beta} 2h\,
{\overline{\cal M}}^{hh'}_{\alpha\beta}(\p_2,\underline{p_1})^\dagger {\overline{\cal M}}^{hh'}_{\alpha\beta}(\p_2,\underline{p_1}) \bigg|_{p_1^+=p_2^+} \, .
\ea
After inserting the antiquark scattering amplitude, in the definition of the antiquark helicity asymmetry, one can easily arrive at
\ba
\label{eq:cross-section-pol-explicit-antiquark}
\frac{d^2\Delta\sigma^{ \bar{q}A\to  \bar{q}+X}}{d^2\p_2}
&=& \frac{1}{N_c} \frac{1}{(2\pi)^2}
\int d^2 \z' \int d^2 \z\,  e^{-i(\p_2-\p_1)\cdot(\z\!-\!\z')} \, \frac{1}{4p_1^+}\\
&&
\hspace{-1.2cm}
\times \, {\rm tr} \,
\bigg\{\int_{-\frac{L^+}{2}}^{\frac{L^+}{2}}dz^+\,
\mathcal{U}_F(\z' )\,
\mathcal{U}_F^\dagger \Big(z^+,-\frac{L^+}{2};\z\Big)\,
\big[\epsilon^{ij}\left( i g t\! \cdot \! \mathcal{F}_{ij}(\uz) \right)\big]
\mathcal{U}^\dagger_F\Big(\frac{L^+}{2},z^+;\z\Big)
\nn\\
&&
\hspace{-0.9cm}
+\int_{-\frac{L^+}{2}}^{\frac{L^+}{2}}dz^+\,
\mathcal{U}_F\Big(\frac{L^+}{2},z^+;\z'\Big)\,
\big[\epsilon^{ij}\left( -i g t\! \cdot \! \mathcal{F}_{ij}(z^+,\z') \right)\big]
\mathcal{U}_F \Big(z^+,-\frac{L^+}{2};\z'\Big)
\mathcal{U}^\dagger_F (\z)
\Bigg\} \, .\nn
\ea
Similar to the procedure followed in the study of the quark helicity asymmetry, we introduce the target averaging of the antiquark helicity asymmetry. Again, using the same change of varibales, one can write this expression in terms of $\r$ and $\b$. Finally, after using the definition of the decorated dipole $O_{(3)}$ given in Eq. \eqref{O3}, one simply gets the target-averaged antiquark helicity asymmetry as
\ba
\Big\langle \frac{d^2\Delta\sigma^{ \bar{q}A\to  \bar{q}+X}}{d^2\p_2} \Big\rangle_A =
 \frac{1}{(2\pi)^2}\int d^2\r \, e^{-i(\p_2-\p_1)\cdot\r} \frac{i}{4p_1^+} \Big[ O^\dagger_{(3)}(\r)-O_{(3)}(-\r)\Big] \, .
\ea
After using the symmetry properties of $O_{(3)}$ given in Eq. \eqref{sym-O3}, one arrives to the final expression for the target-averaged antiquark helicity asymmetry given in Eq. \eqref{final-antiquark-helicity}. Again, in order to have an easy comparison of the results, in the final step we have renamed  $\p_2\mapsto\q$ and $\underline{p_1}\mapsto \underline{k}$ when we are writing Eq. \eqref{final-antiquark-helicity}.

\end{document}